%% file: main.tex
\documentclass[preprintnumbers,floatfix,superscriptaddress,nofootinbib,pra,twocolumn,showpacs]{revtex4-2}

\usepackage{graphicx} 
\usepackage{dcolumn} 
\usepackage{bm} 
\usepackage{helvet}
\usepackage{microtype}
\makeatletter
\renewcommand{\figurename}{Figure}
\renewcommand*{\fnum@figure}{{\normalfont\bfseries \figurename~\thefigure}}
\renewcommand*{\@caption@fignum@sep}{\textbf{: }}

\renewcommand{\tablename}{Table}
\renewcommand*{\fnum@table}{{\normalfont\bfseries \tablename~\thetable}}
\makeatother

\usepackage{algorithm}
\usepackage{algpseudocode}

\usepackage{tikz}
\usetikzlibrary{arrows.meta,decorations.pathmorphing,backgrounds,positioning,fit,petri}
\definecolor{tudelft-sea-green}{cmyk}{0.54,0,0.32,0}
\definecolor{tudelft-green}{cmyk}{1,0.15,0.4,0}
\definecolor{tudelft-dark-blue}{cmyk}{1,0.66,0,0.4}
\definecolor{tudelft-purple}{cmyk}{0.98,1,0,0.35}
\definecolor{tudelft-turquoise}{cmyk}{0.82,0,0.21,0.08}
\definecolor{tudelft-sky-blue}{cmyk}{0.45,0,0.06,0.06}
\definecolor{tudelft-lavendel}{cmyk}{0.45,0.2,0,0.07}
\definecolor{tudelft-warm-purple}{cmyk}{0.58,1,0,0.02}
\definecolor{tudelft-fuchsia}{cmyk}{0.19,1,0,0.19}

\usepackage{hyperref}
\hypersetup{
    colorlinks=true,
    citecolor=tudelft-green,
    linkcolor=tudelft-purple,
    urlcolor=gray
}

\usepackage{mathtools}
\usepackage{amssymb}
\usepackage{amsthm}
\usepackage{bbm}  

\usepackage{braket}
\usepackage{mleftright}  

\usepackage{xcolor}

\usepackage[shortlabels]{enumitem}
\usepackage{booktabs}

\DeclarePairedDelimiter\abs{\lvert}{\rvert}%
\DeclarePairedDelimiter\norm{\lVert}{\rVert}%

\makeatletter
\let\oldabs\abs
\def\abs{\@ifstar{\oldabs}{\oldabs*}}
\let\oldnorm\norm
\def\norm{\@ifstar{\oldnorm}{\oldnorm*}}
\makeatother

\renewcommand\bra[1]{{\langle{#1}|}}
\renewcommand\ket[1]{{|{#1}\rangle}}

\newcommand*\mean[1]{\bar{#1}}

\date{\today}

\newcommand{\ki}{k_i}
\newcommand{\vi}{v_i}
\newcommand{\ps}{p_{\mathrm{swap}}}
\newcommand{\tcut}{t_{\mathrm{cut}}}
\newcommand{\Fgen}{F_{\mathrm{new}}}
\newcommand{\Fmin}{F_{\mathrm{min}}}
\newcommand{\pgen}{p_{\mathrm{gen}}}
\newcommand{\di}{d}

\theoremstyle{definition}
\newtheorem{definition}{Definition}

\begin{document}

\title{Continuously Distributing Entanglement \\ in Quantum Networks with Regular Topologies}

\title[Continuously Distributing Entanglement in Quantum Networks with Regular Topologies]{\texorpdfstring{Continuously Distributing Entanglement \\ in Quantum Networks with Regular Topologies}{Continuously Distributing Entanglement in Quantum Networks with Regular Topologies}}

\author{Lars Talsma}
 \affiliation{QuTech, Delft University of Technology, Lorentzweg 1, 2628 CJ Delft, The Netherlands}

\author{Álvaro G. Iñesta}
 \affiliation{QuTech, Delft University of Technology, Lorentzweg 1, 2628 CJ Delft, The Netherlands}
 \affiliation{EEMCS, Quantum Computer Science, Delft University of Technology, Mekelweg 4, 2628 CD Delft, The Netherlands}
 \affiliation{Kavli Institute of Nanoscience, Delft University of Technology, Lorentzweg 1, 2628 CJ Delft, The Netherlands}
 
\author{Stephanie Wehner}
 \email{s.d.c.wehner@tudelft.nl}
 \affiliation{QuTech, Delft University of Technology, Lorentzweg 1, 2628 CJ Delft, The Netherlands}
 \affiliation{EEMCS, Quantum Computer Science, Delft University of Technology, Mekelweg 4, 2628 CD Delft, The Netherlands}
 \affiliation{Kavli Institute of Nanoscience, Delft University of Technology, Lorentzweg 1, 2628 CJ Delft, The Netherlands}

\begin{abstract}
    \noindent Small interconnected quantum processors can collaborate to tackle quantum computational problems that typically demand more capable devices. These linked processors, referred to as quantum nodes, can use shared entangled states to execute nonlocal operations. As a consequence, understanding how to distribute entangled states among nodes is essential for developing hardware and software. We analyze a protocol where entanglement is continuously distributed among nodes that are physically arranged in a regular pattern: a chain, a honeycomb lattice, a square grid, and a triangular lattice. These regular patterns allow for the modular expansion of networks for large-scale distributed quantum computing. Within the distribution protocol, we investigate how nodes can optimize the frequency of attempting entanglement swaps, trading off multiple entangled states shared with neighboring nodes for fewer states shared with non-neighboring nodes. We evaluate the protocol's performance using the virtual neighborhood size---a metric indicating the number of other nodes with which a given node shares entangled states. Employing numerical methods, we find that nodes must perform more swaps to maximize the virtual neighborhood size when coherence times are short. In a chain network, the virtual neighborhood size's dependence on swap attempt frequency differs for each node based on its distance from the end of the chain. Conversely, all nodes in the square grid exhibit a qualitatively similar dependence of the virtual neighborhood size on the swap frequency.
\end{abstract}

\maketitle

\input{introduction}
\input{network-model}
\input{results}
\input{discussion}

\bibliography{bib}

\input{acknowledgements}
\input{author-contributions}

\begin{appendix}
    \input{network-model-details}
    \input{sampling}
    \input{extended-data}
\end{appendix}

\end{document}

%% file: introduction.tex
\section{Introduction}
\noindent A quantum network is a system of interconnected quantum devices that extends beyond the capabilities of classical networks~\cite{kimble2008quantum}. Such devices, known as quantum \emph{nodes}, can be connected over long and short distances. Leveraging quantum-mechanical effects like entanglement, these nodes enable a variety of technologies. For instance, the quantum internet aims to facilitate quantum communication between any two points on Earth~\cite{wehner2018quantum, RFC9340}, allowing applications such as quantum key distribution~\cite{ekert1991quantum,bennett2014quantum} and secure access to remote quantum computers~\cite{broadbent2009universal}. Over short distances, dense arrays of closely connected nodes can cooperate to solve challenging quantum computational problems by distributing the workload among them~\cite{jiang2007distributed, nickerson2014freely}.

\begin{figure}[b]
\includegraphics{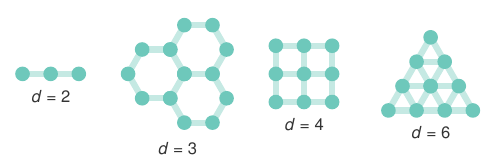}
\caption{\label{fig:topology}\textbf{Quantum networks with a regular topology}. Regularly spaced quantum nodes connected over identical physical channels each have $\di$ physical neighbors and form a chain, a honeycomb lattice, a square grid, and a triangular lattice, as exemplified by these fragments of larger networks.}
\end{figure}

In a network of nodes connected over short distances, we assume that we can design the topology, unlike in long-distance networks whose topologies might have logistical constraints such as the location of cities. Specifically, we investigate network topologies where nodes form a regular pattern, which constitutes a modular and scalable architecture. In particular, we consider networks with a \emph{regular topology} where nodes are regularly spaced and connected over the same physical distance, each having the same number of physical neighbors $\di$. The nodes can form a chain for $\di=2$, a honeycomb lattice for $\di=3$, a square lattice for $\di=4$, and a triangular lattice for $\di=6$~(Figure~\ref{fig:topology}). These lattices regularly tile the plane. 

Such a modular network design offers a scalable approach to, e.g., distributed quantum computing. In such a setting, nodes can implement nonlocal operations using entangled states shared with other nodes, allowing for universal quantum computation~\cite{gottesman1999demonstrating, cirac1999distributed, eisert2000optimal}. Consequently, nodes can scale the number of qubits available for computation by sharing such bipartite states---\emph{entangled links}---with many different nodes. Additionally, since entangled links are consumed in nonlocal operations, nodes require many links if they want to implement many nonlocal operations. 

To distribute many links among many different nodes, we consider a protocol for \emph{continuous distribution}~(CD) of entanglement, where nodes continuously distribute links among them~\cite{inesta2023performance}. Compared to \emph{on-demand} protocols where nodes explicitly request entanglement~\cite{chakraborty2019distributed, vardoyan2019stochastic, inesta2023optimal}, CD protocols do not involve a routing problem to establish links among the nodes requesting them~\cite{chakraborty2019distributed, Pant2019, inesta2023performance}. Solving routing problems can be computationally demanding for large quantum networks. Hence, in networks with many nodes, employing a CD protocol could be a suitable approach to distributing entangled links at a high rate. 

We investigate the performance of a CD protocol that aims to distribute many entangled links among many different nodes in quantum networks with a regular topology, employing the quantum network model of Ref.~\cite{inesta2023performance}. In such networks, pairs of nodes are physical neighbors when they are connected by physical channels such as optical fibers~\cite{tittel1998violation}. Nodes can store quantum information in the form of qubits, and nodes can entangle these qubits to form entangled links. We assume that nodes can share multiple links and that each node has a ``large-enough'' number of quantum memories with finite coherence times (see Section~\ref{subsec:inf_networks} for details). We assume that all nodes are identical and that they are connected over identical physical channels (in particular, all neighboring nodes are at the same physical distance). Moreover, each node has the same number of physical neighbors in the resulting networks. Other unit cells can tile the plane (see Figure~2(a) in Ref.~\cite{harney2022analytical}), although these generally involve nodes with different numbers of physical neighbors or physical channels with different lengths. Therefore, we restrict ourselves to the regular patterns from Figure~\ref{fig:topology}. Lastly, we consider networks with and without boundaries---\emph{finite} and \emph{infinite} networks.

To distribute useful entanglement among the networked nodes, ($i$) physical neighbors can generate shared entangled links in a heralded fashion~\cite{bernien2013heralded}, ($ii$) remote nodes can transform two links shared with an intermediary node into a longer link in an entanglement swap~\cite{zukowski1993event}, and ($iii$) nodes can discard links when their quality has decreased too much due to decoherence and entanglement swaps~\cite{collins2007multiplexed, rozpkedek2018parameter, inesta2023performance}. Following Ref.~\cite[Algorithm~1]{inesta2023performance}, we encapsulate these operations in a simple CD protocol that discretizes time and prescribes the nodes what to do in each time step. Within the protocol, the nodes can adjust how frequently they attempt swaps to modify the distribution of entangled links: if no swaps are performed, entanglement will only be shared among physical neighbors; if swaps are often performed, entanglement will mostly be shared among physically distant nodes.

With the objective of distributing many entangled links among many different nodes, we evaluate the performance of the CD protocol using the performance metrics introduced in Ref.~\cite{inesta2023performance}. At a specific time, the \emph{virtual neighborhood size} indicates the number of nodes any node shares entangled links with, and the \emph{virtual node degree} indicates how many entangled links any node shares with other nodes. The virtual neighborhood size and virtual node degree explicitly consider the time dependence of entangled links as links can be created and removed over time. For example, in a distributed quantum computing setting, a large virtual neighborhood size means that a node can perform nonlocal operations with many different nodes, and a large virtual node degree means that a node can implement many nonlocal operations. Other approaches to analyzing entanglement distribution in quantum networks include evaluating the time it takes to distribute end-to-end entanglement among specific pairs of nodes~\cite{azuma2021tools, inesta2023optimal, khatri2019practical, shchukin2019waiting, shchukin2022optimal}. However, such a metric is better suited to evaluate the performance of on-demand protocols where the goal is to optimize the time it takes to generate entanglement among a set of end nodes. The virtual neighborhood size and virtual node degree are more suitable for evaluating CD protocols, for example capturing the goal of distributing many links among many different nodes. 

In this paper, we employ numerical methods to evaluate the performance of a CD protocol distributing entanglement in regular-topology quantum networks to maximize the virtual neighborhood size and virtual node degree. Our main findings offer design heuristics for CD protocols in quantum networks with regular topologies:
\begin{itemize}[leftmargin=*]
    \item When coherence times are short, swaps must be performed more frequently to maximize the virtual neighborhood size. Intuitively, nodes must make good use of the links before the links are cut off.
    \item The impact of network boundaries on the protocol's performance depends on the network topology. In a finite chain, the dependence of the virtual neighborhood size on the swap attempt frequency is different for each node depending on the node's distance to the edge of the chain. In contrast, for networks with a square-lattice topology, the virtual neighborhood size of all nodes behaves similarly as a function of the swap frequency.
\end{itemize}
\noindent This paper is structured as follows. In Section~\ref{sec:network-model}, we present the quantum network model by discussing networks with a regular topology and how nodes can distribute entanglement in such networks. Furthermore, we adopt a simple CD protocol to facilitate entanglement distribution and define the performance metrics. Subsequently, in Section~\ref{sec:results}, we use these metrics to evaluate the performance of the CD protocol in quantum networks with a regular topology. Finally, in Section~\ref{sec:discussion}, we reflect on the results and discuss potential future work.

%% file: network-model.tex
\section{Quantum network model} \label{sec:network-model}
\noindent We introduce our quantum network model in this section. We first present the physical topologies of the networks we have investigated. Then, we discuss how quantum nodes can use entanglement generation, entanglement swaps, and the removal of low-fidelity links to distribute useful entanglement in regular networks. Accordingly, we adopt an entanglement distribution protocol that the nodes use for distributing entanglement among them. Finally, we discuss the performance metrics that we have used to evaluate the performance of this protocol.

We adopt the quantum network model of Ref.~\cite{inesta2023performance} (see Figure~\ref{fig:example_network} for an illustration). In this model, \emph{nodes} can generate, process, and store quantum information in the form of \emph{qubits}. Such nodes can send quantum information to each other over \emph{physical channels}. Nodes connected via physical channels are physical neighbors. Two nodes can share any number of entangled qubits, where we refer to these shared bipartite states as \emph{entangled links}. Nodes can employ qubit platforms such as nitrogen-vacancy centers in diamond~\cite{bernien2013heralded, hensen2015loophole} and trapped ions~\cite{moehring2007entanglement, stephenson2020high} and be connected over physical channels such as optical fibers~\cite{tittel1998violation} and free space~\cite{hughes2002practical, yin2017satellite}.

\tikzstyle{node}=[draw,circle,tudelft-sea-green,fill=tudelft-sea-green,text=black,minimum width=10mm]
\tikzstyle{ghost_node}=[draw,circle,tudelft-sea-green,fill=tudelft-sea-green,text=black,minimum width=2.5pt]
\tikzstyle{qubit}=[draw,circle,tudelft-green,fill=tudelft-green,text=black,minimum width=1.5mm, inner sep=0pt]
\tikzstyle{ghost_qubit}=[draw,circle,tudelft-green,fill=tudelft-green,text=black,minimum width=1mm, inner sep=0pt]

\def\height{0.707}
\def\width{0.707}
\def\scale{0.6}
                           
\begin{figure}[ht]
\centering
{\footnotesize
    \begin{tikzpicture}[auto, thick, scale=0.5]
  \node at (4.95, 2) [label=right:{\color{black!80}\textsf{Physical channel}}] {};
  \draw [gray!80, line width=1pt] (4.30, 2) -- (5.2, 2);
  
  \foreach \place/\name in {{(0,0)/a}, {(4,0)/b}, {(4,4)/c}}
        \node[ghost_node] (\name) at \place {};

    \foreach \source/\dest in {a/b, b/c}
        \path[tudelft-green!20, line width=5mm] (\source) edge (\dest);
    
    \node at (4.95, 4.05) [label=right:{\color{black!80}\textsf{{Node}}}] {};
    \draw [gray!80, line width=1pt] (4.7, 4) -- (5.2, 4);
    
    \foreach \place/\name in {{(0,0)/a}, {(4,0)/b}, {(4,4)/c}}
        \node[node] (\name) at \place {};

    \node at (4.95, 1.4) [label=right:{\color{black!80}\textsf{Entangled link}}] {};
    \draw [gray!80, line width=1pt] (4.17, 1.4) -- (5.2, 1.4);
        
    \foreach \place/\name in {{(0+1*\scale,0)/a}, {(0+\width*\scale,0+\width*\scale)/aa},  {(0,0+1*\scale)/aaa}, {(4-1*\scale,0)/b}, {(4,0+1*\scale)/bbb}, {(4,4-1*\scale)/c}, {(4-\width*\scale,4-\width*\scale)/cc}, {(4-+1*\scale,4)/ccc}}
        \node[ghost_qubit] (\name) at \place {};
        
    \foreach \source/\dest in {aa/cc, bbb/c}
        \draw[tudelft-warm-purple!60, line width=1.5pt, decorate, decoration={snake, post length=0.5mm, pre length=0.2mm, segment length=4mm, amplitude=1mm}] (\source) -- (\dest);
    
    \draw[tudelft-warm-purple!60, line width=1.5pt, decorate, decoration={snake, post length=0.5mm, pre length=0.75mm, segment length=4mm, amplitude=1mm}] (ccc) arc (111:160:5.9);    
    
    \node at (-3, 0) {}; 

    \node at (4.95, 3.4) [label=right:{\color{black!80}\textsf{Qubit}}] {};
    \draw [gray!80, line width=1pt] (4.1,3.4) -- (5.2, 3.4);

    \foreach \place/\name in {{(0+1*\scale,0)/a}, {(0+\width*\scale,0+\width*\scale)/aa},  {(0,0+1*\scale)/aaa}, {(4-1*\scale,0)/b}, {(4,0+1*\scale)/bbb}, {(4,4-1*\scale)/c}, {(4-\width*\scale,4-\width*\scale)/cc}, {(4-+1*\scale,4)/ccc}}
        \node[qubit] (\name) at \place {};    


\end{tikzpicture}
    \caption{\textbf{Example quantum network.} Nodes can share any number of entangled links, either directly generating entanglement over physical channels or swapping entanglement to create longer-distance links.}
    \label{fig:example_network}
}
\end{figure}
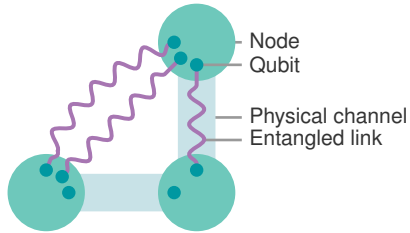

\subsection{Network topology}
\noindent We consider arrays of nodes connected over short distances for large-scale quantum networks. In this scenario, we assume that we are free to design the network topology (in contrast to long-distance networks for quantum communication whose topologies might have logistical constraints such as the location of cities or are utilizing existing optical fibers). Hence, to modularly scale the network size, we investigate network topologies where nodes form a regular pattern.



In particular, we consider quantum networks with a \emph{regular topology} where nodes are regularly spaced and connected over identical physical channels such that each node has the same number $\di$ of physical neighbors. We say that such a network has a \emph{physical node degree} $\di$. The quantum nodes form a chain for $\di=2$, a honeycomb lattice for $\di=3$, a square lattice for $\di=4$, and a triangular lattice for $\di=6$ (Figure~\ref{fig:topology}). The honeycomb, square, and triangular lattices tile the plane regularly. 


We investigate \emph{infinite} and \emph{finite} quantum networks, that is, networks without and with boundaries. The virtual neighborhood size and virtual node degree of nodes in an infinite network approximate, on average, those of nodes far from the network boundaries in large-scale quantum networks (see Section~\ref{subsec:fin_networks} for details on how ``far''). Furthermore, infinite networks provide a convenient platform for performance analysis as each node behaves equivalently due to the network's translational symmetries. We note that nodes on the boundary of finite networks have fewer than $\di$ physical neighbors.


\subsection{Network dynamics} \label{subsec:dynamics}
\noindent To distribute useful entanglement, ($i$) physical neighbors attempt to generate entanglement in a heralded fashion, ($ii$) nodes swap entanglement to convert short-distance entangled links into longer links, and ($iii$) nodes discard low-fidelity entangled links that are of insufficient quality for their intended purpose. We briefly discuss these three operations and summarize the network model parameters in Table~\ref{tab:simulation_parameters} (see Appendix~\ref{app:network-model-details} for more details). 

\paragraph*{\textup{\textbf{Generating entanglement}}} Two physical neighbors can attempt to generate a shared entangled link in a heralded fashion~\cite{bernien2013heralded}. The entanglement generation attempt heralds success with a probability $\pgen$ and fails to generate an entangled link with a probability $1-\pgen$. To model quantum noise, we apply a depolarizing channel (a worst-case noise model)~\cite{nielsen2010quantum} to the Bell state $\ket{\phi^+}=\mleft(\ket{00}+\ket{11}\mright)/\sqrt{2}$. Then, nodes generate entangled links of the Werner form~\cite{werner1989quantum}
\begin{equation} \label{eq:werner}
    \rho = \frac{4F-1}{3}\ket{\phi^+}\bra{\phi^+}+\frac{1-F}{3}\mathbb{I}_4,
\end{equation}
where $F\equiv \Braket{\phi^+ | \rho | \phi^+}$ is the fidelity~\cite{jozsa1994fidelity} of the generated state~$\rho$ to the target state~$\ket{\phi^+}$, and $\mathbb{I}_4$ the four-dimensional identity. We assume that all nodes generate entangled Werner states with the same fidelity $F=\Fgen$.

\paragraph*{\textup{\textbf{Swapping entanglement}}} Two nodes that are not connected by a physical channel can create shared entangled links by swapping entanglement via an intermediary node~\cite{zukowski1993event}. For example, suppose that nodes $\mathrm{A}$ and $\mathrm{B}$ do not share a physical channel. However, nodes $\mathrm{A}$ and $\mathrm{B}$ both share an entangled link with an intermediary node~$\mathrm{I}$ with fidelities $F_{\scriptscriptstyle \mathrm{AI}}$ and $F_{\scriptscriptstyle\mathrm{BI}}$. The intermediary node~$\mathrm{I}$ can do a Bell-state measurement on the qubits storing the links. Then, using local operations and classical communication, the nodes perform an entanglement swap. Specifically, the nodes transform the initial two links into a new link between nodes $\mathrm{A}$ and $\mathrm{B}$ with fidelity $F_{\scriptscriptstyle\mathrm{AB}} \leq F_{\scriptscriptstyle\mathrm{AI}}, F_{\scriptscriptstyle\mathrm{BI}}$~\cite{munro2015inside, inesta2023optimal}. Nodes successfully execute the swap with a probability $\ps$ and fail to generate a longer link with a probability $1-\ps$ (consuming the initial links).

\paragraph*{\textup{\textbf{Discarding entanglement}}} To ensure that entangled links are of sufficient quality for, e.g., distributed quantum computing applications, nodes discard low-fidelity entangled links. We consider two fidelity-decreasing processes. 

Qubits interact with their environment, and the fidelity of the links they store decreases over time---links \emph{decohere}. We assume that the link fidelity decays exponentially with time, where we characterize the decay rate by an abstract coherence time $T$. To ensure that the fidelity of all entangled links exceeds some threshold fidelity $\Fmin$, nodes discard entangled links that are stored longer than a \emph{cutoff time} $\tcut$~\cite{inesta2023optimal, collins2007multiplexed, rozpkedek2018parameter}. In particular, nodes monitor the \emph{age} of the entangled links---the time elapsed since the creation of the link---and subsequently discard links with an age equal to the cutoff time.

The entangled link fidelity generally decreases with the number of swaps it has been involved in. Again, to ensure that the fidelity of all entangled links exceeds some minimum fidelity $\Fmin$, nodes discard entangled links that are the fusion of more than $M$ short-distance links (generated between physical neighbors)~\cite{inesta2023performance}. We refer to $M$ as the \emph{maximum swap distance}.

Combining these requirements, nodes that generate entangled links with a fidelity $\Fgen$ and demand links with a minimum fidelity $\Fmin$ must satisfy the relation~\cite{inesta2023optimal}
\begin{equation} \label{eq:inequality}
    \tcut \leq -T \ln \mleft( \frac{3}{4\Fgen-1}\mleft( \frac{4\Fmin-1}{3}\mright)^{\frac{1}{M}} \mright).
\end{equation}

\subsection{Entanglement distribution protocol}
\noindent To distribute entanglement among the nodes, we employ a simplified version of the continuous distribution (CD) protocol from Ref.~\cite[Algorithm 1]{inesta2023performance}. In our CD protocol (Algorithm~\ref{alg:CD}), physical neighbors generate entangled links in a heralded fashion, non-neighboring nodes swap entanglement via intermediary nodes, and nodes discard links when their fidelity has decreased too much. 

\begin{algorithm}[H]
\caption{Example CD protocol}\label{alg:CD}
\textbf{Inputs:} A quantum network with an arbitrary configuration of entangled links. The network has a regular topology characterized by the physical node degree $\di$. The hardware is described by $\pgen$, $\ps$, $T$ and $\Fgen$, and we choose $\tcut$, $M$ and $\Fmin$ (see Table~\ref{tab:simulation_parameters} for details). Nodes can tune the swap attempt probability $q$ to improve the protocol's performance.

\vspace{2mm}
\textbf{Ouptut:} A quantum network with an updated configuration of entangled links.

\vspace{2mm}
\textbf{Algorithm:}
\begin{enumerate}[leftmargin=*]
    \item \textbf{Cutoff time:} Nodes discard entangled links with ages equal to the cutoff time $\tcut$. Nodes first apply cutoffs to ensure they do not use old links later in the protocol.
    \item \textbf{Entanglement generation:} Physical neighbors attempt to generate entangled links and successfully herald a link of fidelity $\Fgen$ with a probability $\pgen$.
    \item \textbf{Entanglement swapping:} All nodes simultaneously perform the following steps:
    \begin{enumerate}[3.1, leftmargin=*]
        \item Nodes randomly choose a link from their memory.
        \item Nodes choose a second link randomly from the set of links stored in differently-oriented qubits.
        \item Nodes attempt to swap the two entangled links with a probability $q$ and succeed with a probability $\ps$ to create a longer link. When nodes do not attempt to swap, the initial links are not used in further attempts. When the swap fails, nodes discard the initial links. 
    \end{enumerate}
    Nodes repeat steps $3.1$--$3.3$ until no more swaps are possible. Nodes are unaware of the swaps of other nodes.
    \item \textbf{Maximum swap distance:} The nodes communicate which swaps they have attempted. Then, nodes discard entangled links that have been generated from more than $M$ short-distance links (generated between physical neighbors).
\end{enumerate}
\end{algorithm}

In the CD protocol, nodes attempt entanglement swaps with a probability $q$. This probability is the only protocol parameter that nodes can ``tune'' to improve the performance of the entanglement distribution process. During each iteration of the CD protocol, nodes attempt swaps until no more swaps are possible.

We assume that nodes distribute pre-shared entanglement, so we omit the consumption of links in applications (in contrast to the CD protocol of Ref.~\cite[Algorithm~1]{inesta2023performance}). The CD protocol discretizes time and defines the operations that all nodes perform simultaneously. The coherence time $T$ and cutoff time $\tcut$ are expressed in units of this discretized time. Furthermore, we assume that each qubit can only generate entanglement with a fixed neighboring node. Then, nodes label the qubit \emph{orientation} as the direction of the physical neighbor. Nodes only swap entangled links from qubits with different orientations to avoid ``unnecessary'' swaps. In a chain, for example, a node only swaps pairs of links where one link is stored in a left-oriented qubit and the other in a right-oriented qubit. This prevents nodes from creating links between two qubits in the same node (for example, preventing a node from swapping two links with their left physical neighbor) or links that could have been generated directly via a physical channel. As all nodes implement the protocol simultaneously, there is no time to communicate and coordinate more elaborate swap strategies. Lastly, the protocol does not consider entanglement distillation (it could be included by modifying the network parameters; see Appendix~\ref{app:network-model-details}). 

\begin{table}[ht]
    \centering
        \footnotesize{
          \caption{\textbf{Quantum network model parameters.} The parameters $T$, $\tcut$, $\Fgen$, $M$ and $\Fmin$ must satisfy~\eqref{eq:inequality}.}
          \label{tab:simulation_parameters}
          \begin{tabular}{@{}ll@{}}
            \addlinespace[0.4em]
            \multicolumn{2}{c}{\textbf{Physical topology}} \\
            \midrule 
            $\di$ & Physical node degree, number of physical neighbors\\
            \addlinespace[0.4em]
            \multicolumn{2}{c}{\textbf{Hardware}}\\
            \midrule
            $\pgen$ & Probability of successfully heralding entanglement\\
            \addlinespace[0.2em]
            $\ps$ & Probability of successfully swapping entanglement\\
            \addlinespace[0.2em]
            $T$ & Coherence time (exponential decay rate of fidelity)\\
            \addlinespace[0.2em]
            $\Fgen$ & Entanglement generation fidelity \\
            \addlinespace[0.4em]
            \multicolumn{2}{c}{\textbf{Software}} \\
            \midrule
            $\tcut$ & Cutoff time\\
            \addlinespace[0.2em]
            $M$ & Maximum swap distance\\
            \addlinespace[0.2em]
            $\Fmin$ & Minimum required entangled link fidelity \\
            \addlinespace[0.4em]
            \multicolumn{2}{c}{\textbf{Protocol}} \\
            \midrule
            $q$ & Probability of attempting an entanglement swap\\
          \end{tabular}
        }
\end{table}

\subsection{Performance metrics}
\noindent We evaluate the performance of the CD protocol (Algorithm~\ref{alg:CD}) in quantum networks with regular topologies. In applications such as distributed quantum computing, the quantum nodes would likely benefit from ($i$) entangled links with many different nodes to scale the number of qubits available for computation and ($ii$) many entangled links with other nodes to implement many nonlocal operations. With these objectives, we employ performance metrics for quantum networks as introduced by Ref.~\cite{inesta2023performance}:
\begin{definition}[\cite{inesta2023performance}]
     The \emph{virtual neighborhood} of node~$i$ at time~$t$, $V_i(t)$, is the set of nodes that share an entangled link with node $i$ at time $t$. Two nodes are \emph{virtual neighbors} if they share at least one entangled link. The \emph{virtual neighborhood size} is defined as $\vi (t) \equiv \abs{V_i(t)}$.
\end{definition}
\begin{definition}[\cite{inesta2023performance}]
    The \emph{virtual node degree} of node $i$ at time $t$, $\ki(t)$, is the number of entangled links connected to node $i$ at time $t$.
\end{definition}
\noindent These performance metrics capture the objective of distributing many entangled links between many nodes, and explicitly incorporate the time-dependent dynamics of quantum networks. In a distributed quantum computing setting, a large $\vi(t)$ indicates that nodes share entangled links with many different nodes (increasing the number of qubits available for computation), and a large $\ki(t)$ means that nodes can implement many nonlocal operations.

The performance metrics are stochastic processes, i.e., (time) sequences of random variables. Considering this time dependence, we investigate the steady-state expected value of the performance metrics to learn more about the long-term behavior of the network. When a quantum network is running the CD protocol of Algorithm~\ref{alg:CD}, Ref.~\cite{inesta2023performance} showed that there is a unique steady-state value for the expected virtual neighborhood size,
\begin{equation} \label{eq:vi}
    \vi \equiv \lim_{t\to\infty} \mathbb{E}\mleft[\vi (t)\mright],
\end{equation}
 and the expected virtual node degree,
\begin{equation}  \label{eq:ki}
    \ki \equiv \lim_{t\to\infty} \mathbb{E}\mleft[\ki (t)\mright].
\end{equation}
We employ discrete-time network simulations that implement the CD protocol (Algorithm~\ref{alg:CD}) to estimate~$\vi$ and~$\ki$; see Appendix~\ref{app:sampling} for details. 

%% file: results.tex
\section[Performance of CD protocols in regular-topology networks]{\texorpdfstring{Performance of CD protocols\\ in regular-topology networks}{Performance of CD protocols in regular-topology networks}}

\label{sec:results}
\noindent In this section, we evaluate the performance of CD protocols in quantum networks with regular topologies. Nodes can optimize the performance of the entanglement distribution protocol by varying the probability of attempting swaps, $q$. In Section~\ref{subsec:inf_networks}, we investigate the influence of network parameters (the coherence time and the entanglement generation fidelity) on the protocol's performance in \emph{infinite} regular networks (i.e., without boundaries). Then, in Section~\ref{subsec:fin_networks}, we investigate the influence of network boundaries on the virtual neighborhood size of (finite) chains and square lattices. 

\subsection{Infinite networks} \label{subsec:inf_networks}
\noindent The behavior of nodes in infinite networks approximates that of nodes far from the network boundaries in large-scale quantum networks. Furthermore, infinite networks present a convenient setting for performance analysis as all nodes behave equivalently due to the network's translational symmetries. We investigate the influence of varying network parameters (the coherence time~$T$ and the entanglement generation fidelity~$\Fgen$) on the performance metrics using infinite networks.

We assume that nodes generate entanglement deterministically ($\pgen = 1$). Although entanglement generation is generally probabilistic, deterministic entanglement distribution protocols can guarantee the delivery of entangled states at specified time intervals~\cite{humphreys2018deterministic, pompili2022experimental}. Deterministic entanglement generation provides a convenient analysis platform as each node generates the same number of entangled links. In Appendix~\ref{app:network-model-details}, we provide a study of the effect of $\pgen<1$ on the protocol's performance. For $\pgen=1, \frac{1}{2}, \frac{1}{4}$, we observe that the optimal swap attempt probability $q$ scales approximately linearly with $\pgen$.


Furthermore, we assume that nodes execute swaps deterministically ($\ps = 1$). Qubit platforms such as nitrogen-vacancy centers in diamond~\cite{pfaff2014unconditional} and trapped ions~\cite{barrett2004deterministic} can realize Bell-state measurements that succeed deterministically to facilitate the entanglement swaps. Deterministic swaps are convenient for analysis as we do not have to consider failed swaps. When swaps do fail, the virtual neighborhood size and the virtual node degree decrease (see Appendix~\ref{app:network-model-details}, Figure~\ref{fig:prob_gen_swap}).


Lastly, we assume that nodes have a ``large-enough'' number of memories. In particular, nodes can store all generated entangled links until they discard them when the links age to the cutoff time; that is, we assume that nodes have at least $\di \cdot \tcut$ qubits. In our numerical experiments, this corresponds to the order of 10--100 memories. Currently, experiments attain memories of, e.g., 10 qubits in diamond nitrogen-vacancy centers~\cite{bradley2019ten}, which we expect to increase in the near future to the values we use in our simulations.

Our network simulations implement the cutoff time~$\tcut$ and the maximum swap distance~$M$. Then, to vary the coherence time~$T$, we find a $\tcut$ that satisfies condition~\eqref{eq:inequality}; similarly, to vary the entanglement generation fidelity~$\Fgen$, we find an $M$ that satisfies~\eqref{eq:inequality}. Illustratively, if the coherence time $T$ is short, nodes discard links quickly (short $\tcut$) regardless of $\Fgen$ and $M$. Similarly, if $\Fgen$ is too low, swapping two links results in a new link with fidelity $F'<\Fmin$, i.e., nodes should not attempt swaps ($M=1$) regardless of $T$ and $\tcut$.

\begin{figure}
\includegraphics{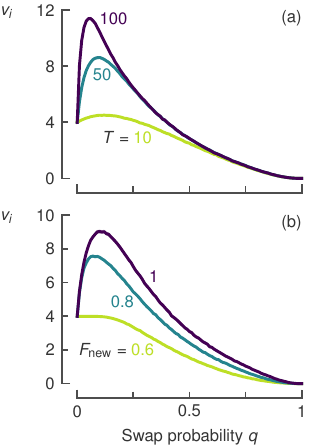}
\caption{\label{fig:inf_sq} \textbf{The optimal probability of attempting to swap~($q$) depends on the coherence time~$(T)$ and entanglement generation fidelity~($\Fgen$)}. Virtual neighborhood size~($\vi$) of a node in an infinite square lattice~($\di=4$) as a function of $q$ for (a) $T=10, 50, 100$ time steps (cutoff time $\tcut=2, 11, 22$ time steps) and (b) $\Fgen=0.6, 0.8, 1$ (maximum swap distance $M=1, 2, 4$); both colored from light to dark. For longer~$T$, links live longer before nodes discard them (longer~$\tcut$), meaning that nodes store more links and can have more virtual neighbors, increasing the maximum~$\vi$. When links live long, swapping conservatively results in more virtual neighbors (lower optimal~$q$). A higher~$\Fgen$ increases the maximum~$\vi$ as links can be involved in more swaps before nodes discard them (increased~$M$). Then, swapping more frequently results in more virtual neighbors (higher optimal~$q$). We consider (a) $\Fgen=0.9$ ($M=3$) and (b) $T=50$ time steps ($\tcut = 11$ time steps). We assume that nodes generate entanglement and execute swaps deterministically ($\pgen, \ps = 1$) and that nodes require a minimum link fidelity $\Fmin = \frac{1}{2}$. Results obtained with network simulations and Monte Carlo sampling with $N=10^4$ realizations per sample, presented with an error band of $\pm 6s/\sqrt{N}$ (generally smaller than the line width), where $s$ is the sample standard deviation.}
\end{figure}

Generally, as a function of increasing swap probability, the virtual neighborhood size starts at $\vi=\di$ ($q=0$). Then, $\vi$ increases to a maximum before converging to zero as $q\to 1$ (Figure~\ref{fig:inf_sq} for a square-lattice network, $\di=4$). For $q=0$, $\vi=\di$ as nodes only share entangled links with their physical neighbors (and discard them when the links reach the cutoff time $\tcut$). Then, as $q$ increases, nodes attempt swaps and can share links with non-neighboring nodes. However, as nodes attempt more swaps (larger~$q$), they also consume more links in swaps and discard more links for being involved in too many swaps ($M$). At first, nodes gain more new virtual neighbors as $q$ increases. Then, at some $q$, losing links (due to a combination of consuming them in swaps and removing low-fidelity ones) balances out this gain in virtual neighbors; $\vi$ reaches a maximum. As the swap probability $q$ increases further, consumption of links in swaps and low-fidelity link removal outweigh the creation of new virtual neighbors, resulting in a decreasing $\vi$. Finally, as $q\to 1$, $\vi\to 0$ as nodes discard all links for being involved in too many swaps.

We see that the swap probability $q$ that maximizes the virtual neighborhood size $\vi$, the \emph{optimal} $q$, depends on the coherence time $T$ and the entanglement generation fidelity $\Fgen$ (Figure~\ref{fig:inf_sq}). As~$T$ increases, entangled links live longer before nodes cut them off (longer $\tcut$). Consequently, nodes store more entangled links and can share entangled links with a larger set of nodes, resulting in an increased~$\vi$. Note that $\vi$ is bounded by a function of the cutoff time~$\tcut$ or the maximum swap distance~$M$ (see Appendix~\ref{app:sampling}, Table~\ref{tab:metric_bounds}). For short $T$ and $\tcut$, nodes quickly discard links for living too long. Then, swapping frequently (relatively high $q$) increases $\vi$ during the short lifetime of the links. In contrast, for relatively long~$T$ and $\tcut$, the nodes benefit from swapping more conservatively as nodes consume fewer links in swaps and discard fewer links that have been swapped too many times. That is, the optimal $q$ decreases for longer $T$. 

As $\Fgen$ increases, entangled links can be involved in more swaps before nodes discard them (larger~$M$). Consequently, nodes can attempt swaps more often (higher~$q$) and share entangled links with nodes that are further away, resulting in an increased virtual neighborhood size~$\vi$. That is, the optimal~$q$ increases with increasing~$\Fgen$. However, increasing~$\Fgen$ results in diminishing gains of the virtual neighborhood size~$\vi$. When there is ``enough'' time for links to exist (large $T$, $\tcut$), the maximum swap distance~$M$ limits~$\vi$. In that case, $\vi$ approaches its upper bound, i.e., nodes share entanglement with all nodes it can potentially share entanglement with. In contrast, when links exist for a limited time, nodes likely do not share entanglement with all nodes they potentially could.

As the physical node degree $\di$ increases, the maximum virtual neighborhood size $\vi$ also increases (see Appendix~\ref{app:extended-data}, Figures~\ref{fig:inf_networks_time_cutoff} and~\ref{fig:inf_networks_max_swap_dist}). In particular, increasing $\di=2$ to $\di=3$ increases the maximum value of $\vi$ by more than the ratio of node degrees ($3/2$). We note that, for increasing swap distance $M$, the bound on $\vi$ grows quicker in networks with $\di=3$ than those with $\di=2$ (see Appendix~\ref{app:sampling}, Table~\ref{tab:metric_bounds}). Increasing the physical node degree to $\di=4,6$ shows diminishing returns.


The virtual node degree $\ki$, i.e., the total number of entangled links connected to node $i$, decreases monotonically for increasing swap probability $q$ for all physical node degrees $\di$ and network parameters (see Appendix~\ref{app:extended-data}, Figures~\ref{fig:inf_networks_time_cutoff} and~\ref{fig:inf_networks_max_swap_dist}). Nodes achieve a maximum $\ki=\di~\cdot~\tcut$ when not attempting swaps ($q=0$) as they only lose links due to cutoffs ($\tcut$). For $q>0$, nodes also consume links in swaps or discard links for being involved in too many swaps. For $q=1$, nodes discard all links for being involved in too many swaps; hence, $\ki= 0$. 


\subsection{Finite networks} \label{subsec:fin_networks}
\begin{figure}
\includegraphics{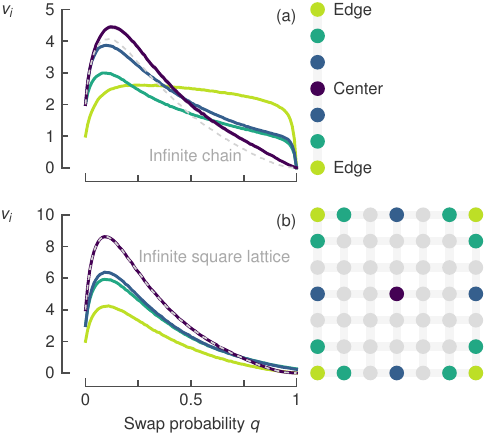}
\caption{\label{fig:fin_networks} \textbf{In a chain, the dependence of the virtual neighborhood size~($\vi$) on the probability of attempting swaps~($q$) is different for each node depending on its distance to the edge; conversely, in a square lattice, the dependence is qualitatively the same across all nodes.} For nodes near and far from the network boundary (colored from light to dark), $\vi$ as a function of~$q$ in a finite (a) chain~($\di=2$) and (b) square lattice~($\di=4$). Edge nodes in a chain only have one physical neighbor and cannot implement swaps. Other nodes can implement swaps but may have a limited potential neighborhood and an asymmetric number of links oriented in either direction, resulting in fewer virtual neighbors. In a finite square lattice, all nodes can implement swaps, and nodes with the same number of physical neighbors have similar~$\vi$. We consider a coherence time $T=50$ time steps (cutoff time~$\tcut=11$ time steps) and an entanglement generation fidelity~$\Fgen=0.9$  (maximum swap distance $M=3$). We assume that nodes generate entanglement and execute swaps deterministically ($\pgen, \ps = 1$) and that nodes require a minimum link fidelity $\Fmin = \frac{1}{2}$. Results obtained with network simulations and Monte Carlo sampling with $N=10^4$ realizations per sample, presented with an error band of $\pm 6s/\sqrt{N}$ (generally smaller than the line width), where $s$ is the sample standard deviation.}
\end{figure}

\noindent We see that the effect of network boundaries on the virtual neighborhood size~$\vi$ depends on the topology~(Figure~\ref{fig:fin_networks}). Specifically, the behavior of $\vi$ depends on the node's distance to the edge of a chain, while, in a square lattice, $\vi$ behaves qualitatively the same across all nodes.

In a finite chain, the virtual neighborhood size $\vi$ of nodes with the same number of physical links to the center node (nodes symmetric around the center) behaves equivalently. Edge nodes only have one physical neighbor, meaning that $\vi=1$ when nodes do not attempt swaps ($q=0$). Furthermore, according to the CD Protocol (Algorithm~\ref{alg:CD}), these edge nodes cannot implement swaps and hence do not consume entangled links in swaps (recall that nodes attempt swaps with entangled links stored in qubits with different orientations, one left-oriented and one right-oriented link in a chain). For $q>0$, $\vi$ of the edge node increases initially as other nodes in the chain attempt swaps. Then, $\vi$ stabilizes for a wide range of swap probabilities before $\vi\to0$ as $q\to 1$ because nodes discard all links for being involved in too many swaps.

Nodes between the edge nodes---\emph{interior} nodes---have virtual neighborhood sizes $\vi$ that are qualitatively more similar to that of nodes in an infinite chain. However, the edges do have an influence. Specifically, compared to nodes in an infinite chain, $\vi$ of interior nodes near the edge converges to zero more slowly as $q\to 1$. We try to explain this behavior with an example. First, note that interior nodes near the edge have an asymmetric number of nodes that they can share entanglement with to their left and right (recall that nodes in a chain swap one left-oriented and one right-oriented link). Now, suppose that all nodes decide to swap entanglement, except for one node that is closer than $M$ physical links to the edge. In that case, this node discards one link that is involved in too many swaps. The other link (oriented toward the nearby edge) cannot have been involved in too many swaps and is not discarded. Hence, at the end of the protocol step, the node near the edge that did not attempt to swap entanglement retains one link, i.e., $\vi\neq0$. Such a scenario is more likely to happen for large~$q$. This results in a larger $\vi$ of nodes near the edge in a finite chain compared to $\vi$ of infinite chain nodes as $q\to 1$. However, when $q=1$, all nodes attempt swaps; then nodes discard all links for being involved in too many swaps ($\vi=0$). Lastly, nodes closer to the edge have fewer nodes they can share entanglement with, resulting in a smaller maximum~$\vi$.

Nodes that are $M$ or more physical links away from the chain edge (e.g., the center node in Figure~\ref{fig:fin_networks}) do have a symmetric number of potential virtual neighbors in both directions. For these nodes, the virtual neighborhood size~$\vi$ behaves similarly as in the case of an infinite chain. Intuitively, such nodes are far enough from the boundary to not experience edge effects. Moreover, nodes that are precisely $M$ physical links away from the edge (the center node in Figure~\ref{fig:fin_networks}) have a higher virtual neighborhood size~$\vi$ compared to nodes in an infinite chain.

In a finite square lattice, the virtual neighborhood size~$\vi$ of nodes with the same physical node degree $d_i$ is quantitatively similar (as boundary nodes in finite regular networks have fewer physical neighbors than interior nodes, we refer to the physical node degree of a specific node~$i$). Corner nodes have two physical neighbors ($d_i=2$), side nodes (on the boundary but not in the corner) have three ($d_i=3$), and interior nodes (not on the boundary) have four ($d_i=4$). In contrast to the finite chain, boundary nodes in a finite square lattice can implement swaps, resulting in a qualitatively similar~$\vi$ behavior of all nodes. However, boundaries do have an influence in that nodes have limited potential virtual neighborhoods when they are closer to the boundary. For instance, side nodes closer to a corner have a smaller~$\vi$ than side nodes further away from a corner. The maximum~$\vi$ of corner nodes is approximately one-half of the maximum~$\vi$ of the interior nodes, and the maximum~$\vi$ of the side nodes is about three-quarters of the maximum~$\vi$ of the interior nodes. Lastly, $\vi$ of the side nodes does not converge to exactly zero as $q\to 1$ because side nodes generate an uneven number ($d_i=3$) of entangled links per time slot, meaning that there is a nonzero probability that the nodes do not involve each link in too many swaps. 

In both finite chains and square lattices, the virtual node degree $\ki$ still monotonically decreases to zero as the swap probability $q\to 1$  (see Appendix~\ref{app:extended-data}, Figure~\ref{fig:fin_networks_inc_ki}). However, when nodes do not attempt swaps ($q=0$), boundary nodes have a smaller virtual node degree, $\ki=d_i\cdot\tcut$. In the finite chain, due to the CD protocol (Algorithm~\ref{alg:CD}), edge nodes do not consume links in swaps because they cannot attempt swaps, meaning that $\ki$ converges to zero slowly. Additionally, by the same explanation as for the virtual neighborhood size~$\vi$, $\ki$ of nodes near the edge decreases to zero slower than nodes in an infinite chain.

%% file: discussion.tex
\section{Discussion} \label{sec:discussion}
\noindent We have adopted a simple protocol for the continuous distribution (CD) of entanglement among networked nodes. The nodes can optimize the probability of attempting entanglement swaps to improve the performance of the CD protocol. Using numerical methods, we have evaluated the protocol's performance in networks where nodes form a regular pattern. We have employed performance metrics that explicitly consider the time dependence of the entangled states. In particular, the virtual neighborhood of any node is the set of nodes it shares entanglement with, and the virtual node degree of any node is the number of entangled states it shares with other nodes. A large virtual neighborhood size indicates that a node shares entangled states with many different nodes, and, for instance, by using entanglement as a means of nonlocal coupling, many qubits are available for computation in distributed quantum computing. In this setting, a sizeable virtual node degree indicates that a node can implement many nonlocal operations. 


We present our findings as heuristics for the design of CD protocols in quantum networks with regular topologies. Firstly, we observe that the swap attempt probability that maximizes the virtual neighborhood size depends on network parameters like the coherence time and entanglement generation fidelity. In particular, this optimal swap attempt probability decreases for a longer coherence time and increases for a higher entanglement generation fidelity. Secondly, the maximum virtual neighborhood size increases with the number of physical neighbors~$\di$ per node---for example, going from a chain~($\di=2$) to a honeycomb lattice~($\di=3$) increases this maximum by more than the ratio of physical neighbors per node~(3/2). Expanding the network to square-lattice~($\di=4$) and triangular-lattice~($\di=6$) topologies shows diminishing returns. Lastly, we see that the influence of network boundaries depends on the network topology. In a chain of nodes, being on or near a network boundary fundamentally alters the performance metrics. In contrast, in a square lattice, the metrics behave qualitatively the same across all nodes. Moreover, the performance metrics of nodes in a finite chain only start approaching those of nodes in an infinite chain when they are further from the boundary than the maximum swap distance $M$.


We have limited our analysis to regular topologies in one and two dimensions, but we could extend it to three-dimensional regular networks, optimally filling rooms with hypothetical future clusters of quantum computing nodes. We have investigated the optimal swap attempt probability while relating the cutoff time with the coherence time and the maximum swap distance with the entanglement generation fidelity; given a setup with a specific coherence time and entanglement generation fidelity, future research could simultaneously optimize over the swap probability, cutoff time, and maximum swap distance to maximize the virtual neighborhood size. Furthermore, our CD protocol delivers pre-shared entangled links. That is, our analysis omitted the consumption of entangled links associated with, for example, implementing nonlocal operations in distributed quantum computing. Such consumption could alter the optimal swap probability and the maximum virtual neighborhood size. Future work could implement link consumption like the analysis of a network with a tree topology from Ref.~\cite{inesta2023performance}.


Further analysis could involve more elaborate CD protocols. For example, the protocol could pair entangled links used in swaps more efficiently (instead of randomly) to increase the virtual neighborhood size, potentially using information about the topology of the network. We assumed that qubits have a ``large-enough'' number of quantum memories; if that assumption is not met, it would be interesting to investigate how a protocol could optimally utilize the limited number of memories. Lastly, protocols could subject the performance metrics to certain constraints. For instance, nodes could demand some minimum virtual neighborhood size $v_{\textrm{min}}$ to ensure a minimum number of qubits available for computation or demand a minimum virtual node degree $k_{\textrm{min}}$ to ensure that nodes can implement a minimum number of nonlocal operations. Nodes could also require the ratio $\ki/\vi$ to attain some minimum value, resulting in a multi-objective optimization problem (similar to meeting the \emph{quality-of-service} requirements investigated by Ref.~\cite{inesta2023performance}).

\bigskip
The data and the code to generate, process, and plot the data can be found in Ref.~\cite{data, *code}.

%% file: acknowledgements.tex
\section{Acknowledgements}
\noindent We thank Janice van Dam, Francisco Ferreira da Silva, and Bethany Davies for their feedback on this manuscript. LT acknowledges SW's financial support to realize this project. ÁGI acknowledges financial support from the Netherlands Organisation for Scientific Research (NWO/OCW), as part of the Frontiers of Nanoscience program. SW acknowledges support from an NWO VICI grant.

%% file: author-contributions.tex
\section{Author contributions}
\noindent ÁGI defined the project, and LT developed and analyzed the network simulations. LT prepared this manuscript. ÁGI and SW supervised the project and provided active feedback at every stage of the project.

%% file: network-model-details.tex
\clearpage 
\onecolumngrid

\section{Quantum network model details} \label{app:network-model-details}
\noindent In this Appendix, we provide more details on the operations and associated parameters we consider in our quantum network model as discussed in Section~\ref{sec:network-model} and motivate the choices of parameters used in Section~\ref{sec:results}. 


\subsection*{Generating entanglement}
\noindent Two physical neighbors herald the successful generation of entanglement~\cite{bernien2013heralded} with a probability $\pgen$, and raise a failure flag with a probability $1-\pgen$. To model quantum noise, we apply a depolarizing channel (a worst-case noise model)~\cite{nielsen2010quantum} to the Bell state $\ket{\phi^+}=\mleft(\ket{00}+\ket{11}\mright)/\sqrt{2}$. As a result of the depolarizing channel, the initial state $\ket{\phi^+}$ is unaffected with some probability~$x$. However, with a probability $1-x$, the initial state depolarizes to the completely mixed state $\mathbb{I}_4/4$, where $\mathbb{I}_4$ is the four-dimensional identity~\cite{nielsen2010quantum}. Consequently, nodes generate entangled links of the Werner form~\cite{werner1989quantum}
\begin{equation} \tag{\ref{eq:werner} revisited}
    \rho = \frac{4F-1}{3}\ket{\phi^+}\bra{\phi^+}+\frac{1-F}{3}\mathbb{I}_4,
\end{equation}
where $F=F(\rho, \ket{\phi^+})\equiv \Braket{\phi^+|\rho|\phi^+}=\frac{3}{4}x+\frac{1}{4}$ is the fidelity~\cite{jozsa1994fidelity} of the generated state~$\rho$ to the target state~$\ket{\phi^+}$. 

Generally, heralded entanglement generation attempts succeed probabilistically, but protocols can guarantee the generation of entangled links at specified intervals~\cite{humphreys2018deterministic, pompili2022experimental}. For example, experiments successfully generate entanglement with a probability $\pgen\approx 5\times 10^{-5}$ between nitrogen-vacancy (NV) centers in diamond~\cite{pompili2021realization, pompili2022experimental}. Such a low probability of generating entanglement would require many entanglement generation attempts per heralded entangled link, resulting in demanding simulation requirements compared to deterministic generation. However, protocols can perform batches of these intrinsically probabilistic entanglement generation attempts to provide deterministic entanglement generation at pre-specified times~\cite{humphreys2018deterministic, pompili2022experimental}. Such protocols can make a trade-off between entanglement generation rates and entanglement generation fidelity, for example, generating entangled links of fidelity $F\approx 0.8$ at a rate of 6 Hz or,  prioritizing the number of entangled links, achieving a generation rate of 39 Hz with $F\approx 0.6$~\cite{humphreys2018deterministic}. Such a robust, deterministic entanglement generation protocol can be part of the link layer in a quantum network stack~\cite{dahlberg2019link, pompili2022experimental}. 

\begin{figure}[ht]
    \centering
    \includegraphics{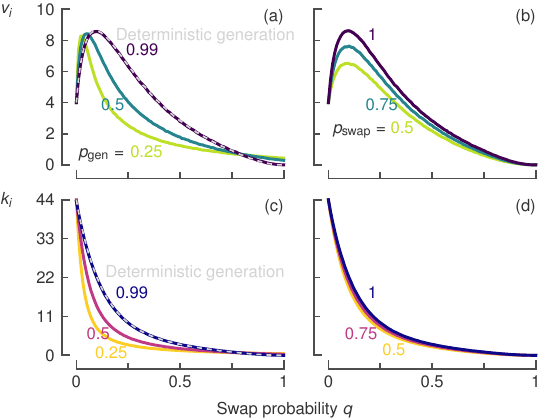}
    \caption{\textbf{Probabilistic generation of entanglement shifts the optimal probability of attempting swaps~($q$) while swaps with nonzero probability of failure decrease the maximum virtual neighborhood~($\vi$)}. (a, b) Virtual neighborhood size~($\vi$) and (c, d) virtual node degree~($\ki$) as a function of~$q$. We vary (a, c) the probability of successfully generating entanglement $\pgen=0.25, 0.5, 0.99$ and (b, d) the probability of successfully executing swaps $\ps=0.5, 0.75, 1$; both colored from light to dark. For the range of $\pgen$ investigated, we observe that the optimal~$q$ scales with $\pgen$ (a). For example, the optimal~$q$ when $\pgen=\frac{1}{2}$ is approximately half of the optimal~$q$ when $\pgen=1$. The maximum~$\ki$ remains approximately the same but converges to zero more quickly for lower $\pgen$ (c). When swaps succeed probabilistically ($\ps<1$), nodes waste links in failed swaps. This means that nodes store fewer links, i.e., lower~$\ki$ (d). Furthermore, nodes have fewer opportunities to share links with remote nodes, resulting in a lower maximum virtual neighborhood size $\vi$ as $\ps$ decreases (b). For probabilistic entanglement generation, we scale the coherence time $T$ (and cutoff time $\tcut$) inversely proportion to $\pgen$ ($T=50, 50, 100, 200$ time steps, $\tcut=11, 11, 22, 44$ time steps for $\pgen=1, 0.99, 0.5, 0.25$). For probabilistic swaps, we consider $T=50$ time steps ($\tcut=11$ time steps). For both cases, we consider an entanglement generation fidelity $\Fgen=0.9$ (maximum swap distance $M=3$) and a threshold fidelity $\Fmin = \frac{1}{2}$. Results obtained using network simulations and Monte Carlo sampling with $N=10^4$ realizations per sample, presented with an error band of $\pm 6s/\sqrt{N}$ (generally smaller than the line width), where $s$ is the sample standard deviation.}
    \label{fig:prob_gen_swap}
\end{figure}

Suppose that we generate entanglement with a low probability $\pgen'$ and that the generated entangled link has a coherence time $T$. Like the entanglement generation protocol discussed above, we use these entanglement generation attempts (with a low probability of success) to create a protocol that generates entanglement with a higher probability of success $\pgen$ over a batch of many individual attempts. Recall that the CD protocol (Algorithm~\ref{alg:CD}) discretizes time in units associated with the rate of the entanglement generation protocol ($\pgen$). That is, as we combine several entanglement generation attempts ($\pgen'$), the entanglement generation protocol ($\pgen$) takes longer and the unit of time increases in the CD protocol. Accordingly, the coherence time $T$ of the entangled link associated with $\pgen$ decreases in units of this discretized time. If we design protocols with $\pgen=1, \frac{1}{2}, \frac{1}{4}$, we observe that the optimal swap attempt probability~$q$ increases approximately linearly with $\pgen$~(Figure~\ref{fig:prob_gen_swap}(a)). As the discretized time step in the CD protocol depends on the time it takes the entanglement generation protocol ($\pgen$) to attempt entanglement delivery between physical neighbors, we scale the coherence time $T$ (and cutoff time $\tcut$) inversely proportional to $\pgen$, e.g., $T(\pgen=1)=2T(\pgen=\frac{1}{2})$. Then, for the range of entanglement generation success probability that we investigated ($\pgen=1, \frac{1}{2}, \frac{1}{4}$), we see that the optimal~$q$ for $\pgen=1$ is approximately double that of the optimal~$q$ when $\pgen=\frac{1}{2}$ and that the optimal~$q$ for $\pgen=\frac{1}{2}$ is approximately double that of the optimal~$q$ when $\pgen=\frac{1}{4}$. 

Lastly, we note that the proof that there exists a unique steady-state value for the expected number of virtual neighbors and the expected virtual degree of any node by Ref.~\cite{inesta2023performance} is under the assumption that entanglement generation is probabilistic ($\pgen<1$). However, in this paper, we generally assume that $\pgen=1$ (deterministic) for a simplified analysis of the results. Reference~\cite{inesta2023performance} expects that there exists a unique steady state for $\pgen=1$; we elaborate on this expectation in Appendix~\ref{app:sampling}. Additionally, we note that the results of the performance metrics are almost indistinguishable for $\pgen=0.99$ and $\pgen=1$ (Figure~\ref{fig:prob_gen_swap}(a,~c)).

\vspace{20mm}

\subsection*{Swapping entanglement}
\begin{figure}[ht]
    \centering
    {\footnotesize
        \begin{tikzpicture}[auto, thick, scale=0.5]



        \foreach \place/\name in {{(0,4)/a}, {(4,4)/b}, {(8,4)/e}}
        \node[ghost_node] (\name) at \place {};

        \foreach \source/\dest in {a/b, b/e}
            \path[tudelft-green!20, line width=5mm] (\source) edge (\dest);
        
        \foreach \place/\name in {{(0,4)/a}, {(4,4)/b}, {(8,4)/e}}
            \node[node] (\name) at \place {};
            
        \foreach \place/\name in {{(0+1*\scale,4)/a}, {(4-1*\scale,4)/b}, {(4+1*\scale,4)/bb}, {(8-1*\scale,4)/e}}
            \node[ghost_qubit] (\name) at \place {};
            
        \foreach \source/\dest in {a/b, bb/e}
            \draw[tudelft-warm-purple!60, line width=2pt, decorate, decoration={snake, post length=0.5mm, pre length=0.75mm, segment length=4mm, amplitude=1mm}] (\source) -- (\dest);
        
        \foreach \place/\name in {{(0+1*\scale,4)/a}, {(4-1*\scale,4)/b}, {(4+1*\scale,4)/bb}, {(8-1*\scale,4)/e}}
            \node[qubit] (\name) at \place {};
        
        \node (arrowstart) at (4, 3) [label=below right:{\color{black!80} \textsf{Swap}}]{};
        \node (arrowend) at (4, 1.75) {}
            edge [latex-, very thick, black!70] (arrowstart);

        \foreach \place/\name in {{(0,0)/a}, {(4,0)/b}, {(8,0)/e}}
            \node[ghost_node] (\name) at \place {};
    
        \foreach \source/\dest in {a/b, b/e}
            \path[tudelft-green!20, line width=5mm] (\source) edge (\dest);
        
        \foreach \place/\name in {{(0,0)/a}, {(4,0)/b}, {(8,0)/e}}
            \node[node] (\name) at \place {};

        \foreach \place/\name in {{(0+1*\scale,0)/a}, {(4-1*\scale,0)/b}, {(4+1*\scale,0)/bb}, {(8-1*\scale,0)/e}}
            \node[ghost_qubit] (\name) at \place {};

        \draw[tudelft-warm-purple!60, line width=2pt, decorate, decoration={snake, post length=0.5mm, pre length=0.4mm, segment length=4mm, amplitude=1mm}] (e) arc (50:130:5.4);    
            
        \foreach \place/\name in {{(0+1*\scale,0)/a}, {(4-1*\scale,0)/b}, {(4+1*\scale,0)/bb}, {(8-1*\scale,0)/e}}
            \node[qubit] (\name) at \place {};
    
    \end{tikzpicture}
        \caption{\textbf{Entanglement swap.} Physical neighbors can directly generate entangled links, while non-neighboring nodes can generate links by swapping entanglement via an intermediary node.}
        \label{fig:swap_example}
    }
\end{figure}
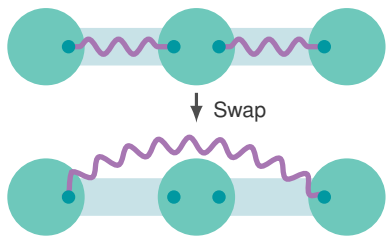

\noindent Two nodes not connected via a physical channel may create an entangled link by swapping entanglement with an intermediary node~\cite{zukowski1993event}. For example, suppose that two nodes $\mathrm{A}$ and $\mathrm{B}$ do not share a physical channel but are both physical neighbors of an intermediary node~$\mathrm{I}$. Nodes $\mathrm{A}$ and $\mathrm{B}$ can directly herald entangled links (of the Werner form~\eqref{eq:werner} with fidelities $F_{\scriptscriptstyle \mathrm{AI}}$ and $F_{\scriptscriptstyle \mathrm{BI}}$) with node~$\mathrm{I}$ over these physical channels. To implement the swap, the intermediary node performs a Bell-state measurement on its entangled qubits. Then, using local operations and classical communication, the three nodes swap entanglement, transforming the initial links into an entangled link between nodes $\mathrm{A}$ and $\mathrm{B}$ of the Werner form with fidelity~\cite{inesta2023optimal, munro2015inside}
\begin{equation}
    F_{\scriptscriptstyle \mathrm{AB}}= F_{\scriptscriptstyle \mathrm{AI}}F_{\scriptscriptstyle \mathrm{BI}}+\frac{(1-F_{\scriptscriptstyle \mathrm{AI}})(1-F_{\scriptscriptstyle \mathrm{BI}})}{3}\leq F_{\scriptscriptstyle \mathrm{AI}}, F_{\scriptscriptstyle \mathrm{BI}}.
\end{equation}
\noindent Swapping entanglement via an intermediary node is illustrated in Figure~\ref{fig:swap_example}. 

We assume that the nodes successfully swap entanglement with a probability $\ps$ and fail with a probability $1-\ps$. On failure, the initial links are destroyed and no link is produced. Experimentally, the Bell-state measurements that facilitate the entanglement swaps can be deterministic ($\ps=1$, for example, in diamond NV centers~\cite{pfaff2014unconditional, pompili2021realization}) or probabilistic (generally $\ps=\frac{1}{2}$ using linear optics~\cite{duan2001long}, but $\ps>\frac{1}{2}$ is possible using ancillary photons~\cite{ewert20143, bayerbach2023bell}). When entanglement swaps succeed probabilistically, nodes can lose entangled links in failed swaps. This means that nodes will store fewer links (decreased~$\ki$) and hence will share links with a smaller set of virtual neighbors, i.e., smaller~$\vi$ (Figure~\ref{fig:prob_gen_swap}(b,~d)). Lastly, we assume that links formed in a swap assume the age of the oldest initial link (the time elapsed since the creation of the link).

\subsection*{Discarding entanglement}
\noindent Qubits interact with their environment and the fidelity of the links they store decreases over time; we say that the links decohere. We model this decoherence as the successive application of a depolarizing channel, a worst-case noise model~\cite{nielsen2010quantum}. Then, during a time interval $\Delta t$, the fidelity $F(t)$ of a Werner state~\eqref{eq:werner} at time $t$ evolves according to~\cite{inesta2023optimal}
\begin{equation}
    F(t+\Delta t)=\frac{1}{4}+\mleft(F(t)-\frac{1}{4}\mright)e^{-\Delta t/T},
\end{equation}
where $T$ is an abstract coherence time that characterizes the exponential decay rate of the fidelity.

To ensure that the fidelity of all entangled links exceeds some threshold fidelity $\Fmin$, nodes discard entangled links that are stored longer than a \emph{cutoff time} $\tcut$~\cite{inesta2023optimal, collins2007multiplexed, rozpkedek2018parameter}. Similarly, nodes discard links that have been formed in the fusion of more than $M$ short-distance links (generated between physical neighbors)---i.e., discard links that exceed the maximum swap distance $M$~\cite{inesta2023performance}. We recall that physical neighbors generate entangled links of the Werner form~\eqref{eq:werner} with a fidelity $\Fgen$ and that new links created in an entanglement swap assume the age of the oldest initial link. Then, given an abstract coherence time $T$ that characterizes the exponential decay rate of the fidelity, nodes must satisfy the relation~\cite{inesta2023optimal}
\begin{equation*} \tag{\ref{eq:inequality} revisited}
    \tcut \leq -T \ln \mleft( \frac{3}{4\Fgen-1}\mleft( \frac{4\Fmin-1}{3}\mright)^{\frac{1}{M}} \mright).
\end{equation*}

Suppose that the entangled link fidelity is not large enough for its intended purpose, e.g., distributed quantum computing applications. In that case, quantum nodes can use entanglement distillation protocols to turn low-fidelity entangled links into links of higher fidelity using local operations. The entangled links that we consider---of the Werner form~\eqref{eq:werner}---are entangled for a fidelity $F>\frac{1}{2}$ ($x>\frac{1}{3}$)~\cite{peres1996separability}. Then, bipartite distillation protocols~\cite{bennett1996purification, deutsch1996quantum} can distill multiple initial entangled links of fidelity $F>\frac{1}{2}$ to a new link of fidelity $F'>F$. In this way, nodes ensuring a minimum fidelity $\Fmin>\frac{1}{2}$ can generate higher-fidelity links if their application requires so. In general, we choose the lower bound $\Fmin=\frac{1}{2}$ to analyze performance at the extreme of ``useful'' links (note that we require $\Fmin=\frac{1}{2}+\epsilon$ for a tiny $\epsilon>0$; however, this $\epsilon$ would have an insignificant influence on calculating Inequality~\eqref{eq:inequality}, so we omit it for simplicity).

Although implementing such distillation is outside the scope of this work, distillation can be incorporated into our model. For example, at the entanglement generation level, we could integrate distillation in the deterministic entanglement generation protocol we discussed above. To account for the time needed to implement the distillation, the unit of discretized time increases and, consequently, $T$ and $\tcut$ decrease in terms of this discretized time. Additionally, we should adjust $\pgen$ and $\Fgen$ according to the results of the distillation protocol. For existing links, we could integrate distillation (as part of an application) in the CD protocol (Algorithm~\ref{alg:CD}) (see also the CD protocol of Ref.~\cite{inesta2023performance} and their discussion of integrating distillation in Appendix~A).

From the parameters related by Inequality~\eqref{eq:inequality}, we adopt the cutoff time $\tcut$ and the maximum swap distance $M$ as simulation parameters. Then, if we want to investigate various coherence times $T$, we associate values of $\tcut$ that satisfy Inequality~\eqref{eq:inequality} and use those values in our simulations. Similarly, if we want to vary the entanglement generation fidelity $\Fgen$, we associate values of $M$ that satisfy Inequality~\eqref{eq:inequality}. In addition to the remarks in Section~\ref{subsec:dynamics}, we require 
\begin{equation} 
    \frac{3}{4\Fgen-1}\mleft( \frac{4\Fmin-1}{3}\mright)^{\frac{1}{M}} < 1 \label{eq:req}
\end{equation}
since the cutoff must be positive.
If we assume $\Fmin=\frac{1}{2}$ and want some $M>1$, we see that there is some minimum~$\Fgen$ to satisfy Inequality~\eqref{eq:req} regardless of the $T$. Lower values of $M$ result in lower required values of $\Fgen$ (assuming constant $\Fmin$) and vice versa, motivating our choice to relate the values of $\Fgen$ and $M$ and those of $T$ and $\tcut$. 

Experimentally, multi-qubit nodes that combine communication and memory qubits can reach coherence times in the order of seconds~\cite{bradley2022robust}. In the CD protocol, we discretize time in units associated with the generation rate of protocols that guarantee the generation of entangled links at specified intervals. As discussed above, such protocols currently deliver (on the order of) tens of links per second. Accordingly, we employ values of the coherence times $T$ that may be feasible in the near future, i.e., coherence times of tens to hundreds of time steps. Using the entanglement generation fidelities we discussed above, we retrieve reasonable values for the maximum swap distance~$M$ via Inequality~\eqref{eq:inequality}.

Lastly, we assume that quantum nodes have a ``large-enough'' number of memories. When not attempting entanglement swaps ($q=0$), nodes can store all entangled links until they discard the links when they age to the cutoff time. This translates into nodes storing at most $\di\cdot \tcut$ entangled links (see Table~\ref{tab:metric_bounds}), where $\di$ is the physical degree of the node. With the moderate values of the cutoff time $\tcut$ we employ in this work, the number of qubits required per node is relatively close to experimentally-achieved values~\cite{bradley2019ten}.

%% file: sampling.tex
\clearpage 
\onecolumngrid

\section{Data sampling} \label{app:sampling}
\noindent In this Appendix, we present our data sampling technique to obtain the performance metrics shown in this manuscript. In particular, to estimate the expected virtual neighborhood size~$\vi$~\eqref{eq:vi} and the expected virtual node degree~$\ki$~\eqref{eq:ki}, we employ discrete-time network simulations that implement the CD protocol (Algorithm~\ref{alg:CD}). We simulate the networks for $\ell$ timesteps and verify (using Algorithm~\ref{alg:steady-state}) that the performance metrics attain their steady state at the end of the simulation during a time window $w$, i.e., at times $t=t_{\ell-w}, t_{\ell-w+1}, \dots, t_{\ell-1}$. If the steady state is achieved, we estimate the expected virtual neighborhood size and expected virtual node degree by sampling the performance metrics at the final timestep over many ($N=10^4$) network simulations,
\begin{align} 
        \vi &\equiv \lim_{t\to\infty} \mathbb{E}\mleft[\vi (t)\mright] \approx \mean{v}_{i, N}(t_{\ell-1}), \label{eq:vi_est} \\
        \ki &\equiv \lim_{t\to\infty} \mathbb{E}\mleft[\ki (t)\mright]\approx \mean{k}_{i, N} (t_{\ell-1}).  \label{eq:ki_est}
\end{align}
Here, $\bar{v}_{i, N}(t_{\ell-1})$ and $\bar{k}_{i, N}(t_{\ell-1})$ are the virtual neighborhood size and virtual node degree of node $i$ at time $t_{\ell-1}$ averaged over a sample with $N$ realizations.

We now discuss the algorithm to find the steady-state expected value of a stochastic process given a sample of~$N$ realizations as introduced by Ref.~\cite[Appendix~D, Algorithm~2]{inesta2023performance}. Specifically, algorithm~\ref{alg:steady-state} determines whether the sample mean of the stochastic process $\{X(t)\}$ over $N$ realizations attains a steady state, which is an adaptation of the steady state algorithm of Ref.~\cite{inesta2023performance} to better to suit our needs. Specifically, we only employ steps 1--4 of the steady state algorithm of Ref.~\cite{inesta2023performance} (further steps determined \emph{when} the steady state starts). Furthermore, compared to the error $\varepsilon$ employed by Ref.~\cite{inesta2023performance}, we adjust the error $\varepsilon'=3\varepsilon$ to ensure that the algorithm declares that the steady state has been reached once we are ``close enough'' to the steady state value (see below for more details).

\begin{algorithm}[H]
\caption{Steady state estimation}\label{alg:steady-state}
\textbf{Inputs:} 
\begin{itemize}
    \item $\mean{X}_N(t)$: the sample mean of the stochastic process $\{X(t)\}$ observed over $N$ realizations at times $t=t_0, t_1, \dots, t_{\ell-1}$;
    \item $a, b$: the minimum and maximum values of the stochastic process $\{X(t)\}$;
    \item $w$: size of the steady-state window.
\end{itemize}

\textbf{Ouptut:} 
\begin{itemize}
    \item Assesment of whether $\mean{X}_N(t)$ has attained a steady state in the time window $W=\{\ell-w, \ell-w+1, \dots, \ell-1\}$.
\end{itemize}

\textbf{Algorithm:}
\begin{algorithmic}[1]
\State Define the error $\varepsilon' \gets 3(b-a)/\sqrt{N}$.
\State Define the steady state window  $W\gets\{\ell-w, \ell-w+1, \dots, \ell-1\}$.
\State Calculate the size of the interval of confidence $\Delta_{ij}\gets 2\varepsilon'-\abs{\mean{X}_N(t_i)-\mean{X}_N(t_j)}, \forall i,j \in W$ and $i\neq j$.
\State If the interval of confidence $\Delta_{ij}<\frac{3}{2}\varepsilon'$, then \textbf{abort} (steady state not found). Otherwise, declare \textbf{success} (steady state present in the time window).
\end{algorithmic}
\end{algorithm}

\noindent We now adjust the results found by Ref.~\cite{inesta2023performance} to conform to the choice of $\varepsilon'$ in Algorithm~\ref{alg:steady-state}. Let us consider a stochastic process $\{X(t)\}$. We assume that $\{X(t)\}$ has a constant, steady state mean, $\lim_{t\to\infty}\mathbb{E}[X(t)]=X_{\infty}<\infty$, and a finite variance, $\sigma(t)^2<\infty$. Observing the stochastic process over $N$ realizations at times $t=\{t_0, t_1,\dots t_{\ell-1}\}$, $t_0<t_1<\dots<t_{\ell-1}$, we denote the value taken in realization $n$ as $x_n(t)$, where $a\leq x_n(t)\leq b$, with $a,b\in\mathbb{R}$. Then, we denote the sample average at time $t$ over $N$ realizations as
\begin{equation}
    \mean{X}_N(t)=\frac{1}{N}\sum_{n=0}^{N-1} x_n(t).
\end{equation}
Now, let us assume that $\{X(t)\}$ has attained a steady state at some time $t=t_\alpha$. Then, considering $X(t_i)$ for all $i\geq \alpha$, we follow Ref.~\cite{inesta2023performance} and use the central limit theorem and the properties of a normal distribution (the probability that a normally distributed random variable takes a value more than six standard deviations from the mean value is approximately $2\times 10^{-9}$), to conclude that 
\begin{equation}\label{eq:prob_ICi}
    \Pr\mleft[\mathbb{E}\mleft[X(t_i)\mright]\in \left(\mean{X}_N(t)-\frac{6\sigma(t)}{\sqrt{N}}, \mean{X}_N(t)+\frac{6\sigma(t)}{\sqrt{N}} \right)\mright]\geq 1-2\times 10^{-9} \approx 1.
\end{equation}
Now, let us define an error $\varepsilon'=3(b-a)/\sqrt{N}$ and consider a confidence interval for the steady-state sample average of $X(t_i)$,
\begin{equation}
    \mathrm{IC}_{i}=\Big(\mean{X}_N(t_i) - \varepsilon', \mean{X}_N(t_i) + \varepsilon'\Big).
\end{equation}
Furthermore, we use that the standard deviation is bounded by $\sigma(t)\leq (b-a)/2$ such that $\varepsilon'=3(b-a)/\sqrt{N}\geq 6\sigma(t)/\sqrt{N}$. Then it follows from probability~\eqref{eq:prob_ICi} that
\begin{equation}
    \Pr\mleft[\mathbb{E}\mleft[X(t_i)\mright]\in \mathrm{IC}_{i}\mright] \approx 1.
\end{equation}

Similarly, we consider the steady-state $X(t_i)$ and $X(t_j)$ for all $i, j\geq \alpha$, and follow Ref.~\cite[Equation~D.9]{inesta2023performance} to find that
\begin{equation}
    \Pr\mleft[\mathbb{E}\mleft[X(t_i)\mright]\in \mathrm{IC}_{ij}\mright]\geq \frac{1-2\times10^{-9}}{2} +\frac{0.9973}{2} \approx 1
\end{equation}
for an interval of confidence $ij$
\begin{equation}
    \mathrm{IC}_{ij}=\Big( \max\mleft(\mean{X}_N(t_i), \mean{X}_N(t_j)\mright) - \varepsilon', \min \mleft(\mean{X}_N(t_i), \mean{X}_N(t_j)\mright) + \varepsilon'\Big).
\end{equation}
This interval indicates the overlap in the intervals of confidence for the steady-state sample averages of $X(t_i)$ and $X(t_j)$ and has a size
\begin{equation}
    \Delta_{ij} = 2\varepsilon'-\abs{\mean{X}_N(t_i)-\mean{X}_N(t_j)}.
\end{equation}
We used that $\varepsilon'=3(b-a)/\sqrt{N}\geq 6\sigma(t)/\sqrt{N}$ and that the probability that a normally distributed random variable takes a value more than six respectively three standard deviations from the mean is approximately $2\times 10^{-9}$ and $0.0027$.

For infinite quantum networks with a regular topology with a physical node degree $\di$, the virtual neighborhood size~$\vi$ and the virtual node degree~$\ki$ are bounded by a function of the cutoff time~$\tcut$ and the maximum swap distance~$M$ (Table~\ref{tab:metric_bounds})~\cite[see Appendix B for details]{talsma2023continuous}. We use these values as the upper bound $b$ in calculating the error $\varepsilon'$. Note that, in finite networks, the boundary nodes have fewer than $d$ physical neighbors such that also the upper bound $b$ is smaller. We also note that the performance metrics are bounded below by $a=0$.

\begin{table}[ht]
    \centering
        \footnotesize{
          \caption{\textbf{Bounds on the virtual neighborhood size $\vi$ and the virtual node degree $\ki$ in infinite regular networks.}}
          \label{tab:metric_bounds}
          \begin{tabular}{@{}lll@{}}
            \toprule 
            \textbf{} &  \multicolumn{1}{c}{$\vi$} &  \multicolumn{1}{c}{$\ki$} \\
            \midrule 
            $\di=2 \quad$ &  $2 \min\big(\tcut, M\big)\quad$ & $2\tcut$ \\
            \addlinespace[0.2em]
            $\di=3 \quad$ &  $3 \min\big(\tcut, \frac{1}{2}M(M+1)\big)\quad$ & $3\tcut$ \\
            \addlinespace[0.2em]
            $\di=4 \quad$ &  $4 \min\big(\tcut, \frac{1}{2}M(M+1)\big)\quad$ & $4\tcut$ \\
            \addlinespace[0.2em]
            $\di=6 \quad$ &  $6 \min\big(\tcut, \frac{1}{2}M(M+1)\big)\quad$ & $6\tcut$ \\
            \bottomrule 
          \end{tabular}
        }
\end{table}  

\begin{figure}[ht]
    \centering
    \includegraphics{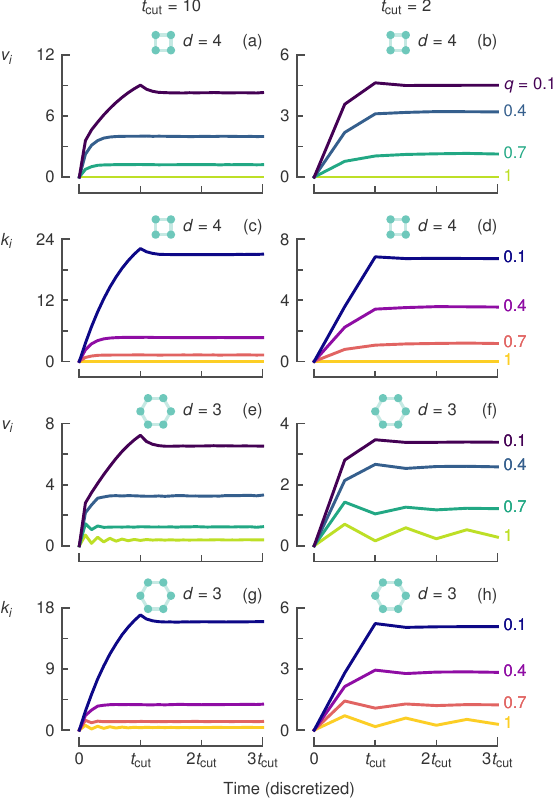}
    \caption{\textbf{Generally, the virtual neighborhood size~($\vi$) and the virtual node degree~($\ki$) quickly converge to a steady state.} Performance metrics (a, c, e, g) $\vi$ and (b, d, f, h) $\ki$ as a function of time in infinite (a--d) square-lattice networks and (e--f) honeycomb-lattice networks. We vary (a, c, e, g) the cutoff time $\tcut=10$ time steps (coherence time $T=45$ time steps) and (b, d, f, h) $\tcut=2$ time steps ($T=9$ time steps) as well as the swap attempt probability $q=1, 0.7, 0.4, 0.1$ (colored from light to dark). In networks with a square-lattice topology~(a--d), the performance metrics reach their steady after approximately $\tcut$ time steps. The situation is similar for infinite honeycomb-lattice networks~(e--h), except that, when nodes attempt swaps often (large~$q$), the performance metrics show periodic oscillations, i.e., the metrics do not attain a steady state. We consider an entanglement generation fidelity $\Fgen=0.9$ (maximum swap distance $M=3$). We assume that nodes generate entanglement and execute swaps deterministically ($\pgen, \ps = 1$) and that nodes require a minimum link fidelity $\Fmin = \frac{1}{2}$. Results obtained using network simulations and Monte Carlo sampling with $N=10^4$ realizations per sample, presented with an error band of $\pm 6s/\sqrt{N}$ (generally smaller than the line width), where $s$ is the sample standard deviation.}
    \label{fig:inf_networks_convergence}
\end{figure}

We execute each network simulation for $3\tcut$ time steps and use a steady state window $w=\tcut$ (except for the finite chain networks (Figures~\ref{fig:fin_networks}(a) and~\ref{fig:fin_networks_inc_ki}(a, c)), which we simulate for $6\tcut$ time steps). Visual inspection (see Figure~\ref{fig:inf_networks_convergence}) showed that the performance metrics converged quickly to their steady states, with Algorithm~\ref{alg:steady-state} confirming that the performance metrics reached the steady state. 

As noted previously by Ref.~\cite{inesta2023performance}, the overlaps between intervals of confidence $\Delta_{ij}$ may be too small in specific scenarios, meaning that Algorithm~\ref{alg:steady-state} aborts, even when a closer visual inspection strongly indicates that the performance metrics attain some form of a steady state. For instance, the error is relatively small when the upper bound $b$ is relatively low, e.g., when the cutoff time $\tcut$ is short, or the maximum swap distance $M$ is low. Then, using the original error $\varepsilon=(b-a)/\sqrt{N}$, the performance metrics (sometimes $\vi (t)$, other times $\ki (t)$) did not attain a steady state for some values of $q$ according to Algorithm~\ref{alg:steady-state}. However, the resulting ``unsteady-state'' values were strongly in line with expectations compared to the steady-state values of nearby swap probabilities $q$. To prevent Algorithm~\ref{alg:steady-state} from aborting in such a situation, we redefine the error $\varepsilon'=3(b-a)/\sqrt{N}=3\varepsilon$ (which is similar to increasing the value of $b$ as proposed by Ref.~\cite{inesta2023performance}). 

We measure the error in the estimate of the expected steady-state values using the standard error $s_{\mean{X}}=s/\sqrt{N}$, where $s$ is the sample ($N$ realizations) standard deviation. The plots in this work show the data as $\mean{X}\pm 6s_{\mean{X}}$, providing a $1-2\times 10^{-9}\approx 100\%$ interval of confidence. Even with such a large confidence interval, most error bands are on the order of or smaller than the plot line width.

\vspace{6mm}

\newpage

\subsection*{Notes on the existence of a unique steady state}
\noindent Reference~\cite{inesta2023performance} showed that there is a unique steady-state value for the expected number of virtual neighbors $\vi\equiv\lim_{t\to\infty} \mathbb{E}\mleft[\vi (t)\mright]$ and expected virtual degree of any node $\ki\equiv\lim_{t\to\infty} \mathbb{E}\mleft[\ki (t)\mright]$ when a quantum network is running the CD protocol of Algorithm~\ref{alg:CD}. This proof is under the assumption that entanglement generation is probabilistic, $\pgen <1$. However, we generally assume $\pgen=1$ to simplify performance analysis. We now elaborate on the assumption that the unique steady state also exists when $\pgen=1$.

The proof of Ref.~\cite{inesta2023performance} uses that the steady-state is unique for an aperiodic, irreducible, positive recurrent Markov chain~\cite[Theorem 9.3.6]{van2014performance}. In particular, they use the ages of all entangled links in the network to represent the state of the network $s$ and the set of all possible states $\mathcal{S}$ (both finite). Then they show that the transition of a state $s(t)$, $t\in\mathcal{N}$ (discrete time steps) does not depend on past information, 
\begin{equation}
    \Pr[s(t+1)=\sigma\mid s(0), s(1),\dots,s(t)]=\Pr[s(t+1)=\sigma\mid s(t)].
\end{equation}
Hence, the state of the network can be modeled as a Markov chain. Then, they show that this Markov chain is aperiodic, irreducible, and positive recurrent using that, for $\pgen<1$, there is a nonzero probability of returning to the initial state (no links). They conclude that the limit $\lim_{t\to\infty} \Pr[s(t)=\sigma]$, $\forall\sigma\in\mathcal{S}$ is unique and exists. Lastly, they express $\vi$ and $\ki$ as a function of this limit to conclude that $\vi$ and $\ki$ also exist and are unique. 

From their simulations, Ref.~\cite{inesta2023performance} also expects a unique steady state for $\pgen=1$. However, they note that ``the main difficulty in proving its existence is that the Markov chain is not always irreducible (the state with no links may not be reachable from some other states since links are generated at the maximum rate).''  We observe almost indistinguishable steady-state behavior when $\pgen=0.99$ compared to $\pgen=1$ (Figure~\ref{fig:prob_gen_swap}(a)). Additionally, we observe that the performance metrics appear to quickly converge to a steady state (Figure~\ref{fig:inf_networks_convergence}). 

We note that for regular networks and a nonzero probability of nodes attempting entanglement swaps ($q>0$), there is a nonzero probability that all links are involved in too many swaps and discarded. That is, the state of the network returns to the initial state of no links (for $q=0$ and $\pgen=1$, the system deterministically reaches $\vi=\di$, $\ki=\di\cdot \tcut$). For example, starting from the initial (no links) state, regular networks with an even number of physical neighbors can pair the even number of generated links ($\pgen=1$) to involve each link in too many swaps (resulting in $\vi, \ki=0$ when $q\neq 0$). However, in a network with a honeycomb topology ($d=3$), nodes generate three links in each time step, meaning that it is more challenging to match all the links in a way where all links are involved in too many swaps. This difficulty in matching links results in nonzero performance metrics in a honeycomb lattice as $q\to 1$~(Figures~\ref{fig:inf_networks_time_cutoff} and~\ref{fig:inf_networks_max_swap_dist}; the performance metrics converge to zero for the other topologies). 

Upon closer inspection (Figure~\ref{fig:inf_networks_convergence}), we see that the performance metrics show periodic oscillations as $q=1$ in a honeycomb lattice network ($d=3$). To illustrate how this happens, assume that nodes will always attempt swaps ($q=1$) and that the performance metrics start without links. Then, nodes generate three links at the first time step and generally swap two links. This makes it challenging to involve each link in too many swaps ($M$), not removing all links at this first time step. Then, in the next time step, nodes again generate three links, meaning some nodes now have four links that nodes can all swap. At the end of this step, nodes have involved more links in too many swaps, thus discarding them, resulting in lower performance metrics. This oscillatory behavior diminishes after some time. For longer cutoff times $\tcut$ and simulation times (recall that we simulate the networks for $3\tcut$ time steps), there is ``enough'' time for this periodicity to vanish (Figure~\ref{fig:inf_networks_convergence}(e,~g)) and for Algorithm~\ref{alg:steady-state} to declare a steady state has been attained. However, for short $\tcut$, the periodicity is still strong after $3\tcut$ time steps and Algorithm~\ref{alg:steady-state} declares there is no steady state (as is the case for $q=0.95, 0.96, \dots, 1$ in Figure~\ref{fig:inf_networks_time_cutoff}(b, f) with $\tcut=2$; we omit those values in the plot for $d=3$). Additionally, the oscillatory behavior persists longer when $\tcut$ is small, diminishing for (significantly) longer simulation times.

Lastly, we note that the proof by Ref.~\cite{inesta2023performance} assumes that the state space of the network is finite. However, Theorem 9.3.6~\cite{van2014performance} also applies to Markov chains with an infinite state space, as is the case for infinite networks. The difficulty becomes showing that the chain is positive recurrent, i.e., for a state to have a finite mean return time (for finite-state space chains, it is sufficient to be irreducible in order to be positive recurrent~\cite[Theorem~9.3.5]{van2014performance}). 

%% file: extended-data.tex
\clearpage 
\onecolumngrid

\section{Extended network simulations} \label{app:extended-data}
\begin{figure}[ht]
    \centering
    \includegraphics{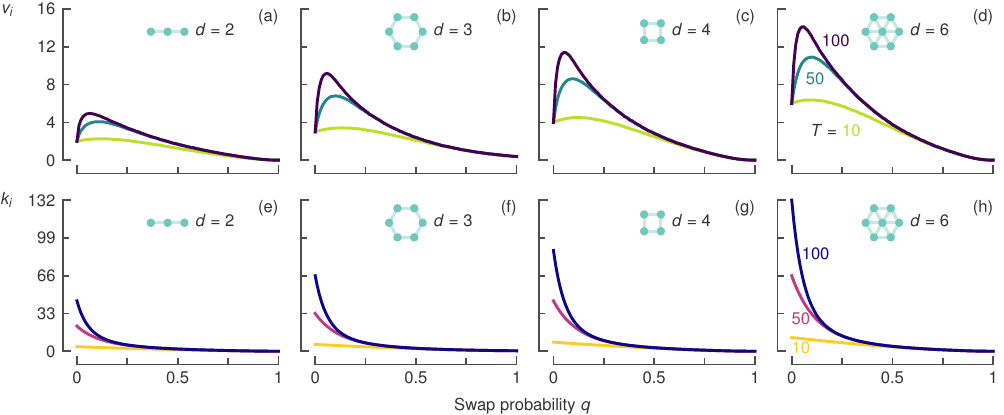}
    \caption{\textbf{The maximum virtual neighborhood size~($\vi$) and virtual node degree~($\ki$) increase with the physical node degree~($\di$).} Performance metrics (a--d) $\vi$ and (e--h) $\ki$ as a function of the swap attempt probability ($q$). We vary the coherence time $T=10, 50, 100$ time steps (cutoff time $\tcut=2,11,22$ time steps; colored from light to dark) in infinite (a, e) chains ($\di=2$), (b, f) honeycomb lattices $(d=3)$, (c, g) square lattices ($\di=4)$ and (d, h) triangular lattices $(d=6)$. The bound on $\vi$ and, consequently, $\vi$ itself increases more than the physical node degree ratio $3/2$ when going from $\di=2$ to $\di=3$ (see Appendix~\ref{app:sampling}, Table~\ref{tab:metric_bounds}). The growth of $\vi$ diminishes when the physical node degree increases further to $\di=4$ and $\di=6$. The number of links connected to node $i$, $\ki$, decreases monotonically from a maximum $\ki=\di\cdot\tcut$ ($q=0$; nodes only share entangled links with physical neighbors and generate entangled links with all their~$\di$ physical neighbors) to $\ki=0$ as $q=1$ (nodes discard all links as nodes involve all of them in too many swaps). We consider an entanglement generation fidelity $\Fgen=0.9$ (maximum swap distance$M=3$). We assume that nodes generate entanglement and execute swaps deterministically ($\pgen, \ps = 1$) and that nodes require a minimum link fidelity $\Fmin = \frac{1}{2}$. Results obtained using network simulations and Monte Carlo sampling with $N=10^4$ realizations per sample, presented with an error band of $\pm 6s/\sqrt{N}$ (generally smaller than the line width), where $s$ is the sample standard deviation. We omit the performance metrics for $\di=3$, $T=10$ time steps, and $q=0.95, 0.96, \dots, 1$ as those simulations did not attain a steady state; see the notes on the existence of a unique steady state in Appendix~\ref{app:sampling}.}
    \label{fig:inf_networks_time_cutoff}
\end{figure}

\begin{figure}[ht]
    \centering
    \includegraphics{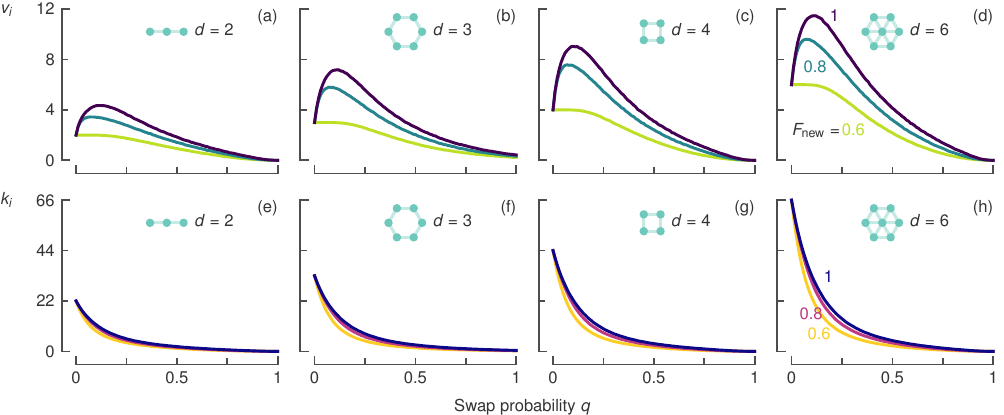}
    \caption{\textbf{The maximum virtual neighborhood size~($\vi$) and virtual node degree~($\ki$) increase with the physical node degree~($\di$).} Performance metrics (a--d) $\vi$ and (e--h) $\ki$ as a function of the swap attempt probability ($q$). We vary the entanglement generation fidelity $\Fgen=0.6, 0.8, 1$ (maximum swap distance $M=1,2,4$; colored from light to dark) in infinite (a, e) chains ($\di=2$), (b, f) honeycomb lattices $(d=3)$, (c, g) square lattices ($\di=4)$ and (d, h) triangular lattices $(d=6)$. We consider a coherence time $T=50$ time steps (cutoff time $\tcut=11$ time steps). We assume that nodes generate entanglement and execute swaps deterministically ($\pgen, \ps = 1$) and that nodes require a minimum link fidelity $\Fmin = \frac{1}{2}$. Results obtained using network simulations and Monte Carlo sampling with $N=10^4$ realizations per sample, presented with an error band of $\pm 6s/\sqrt{N}$ (generally smaller than the line width), where $s$ is the sample standard deviation.}
    \label{fig:inf_networks_max_swap_dist}
\end{figure}

\begin{figure}[H]
    \centering
    \includegraphics{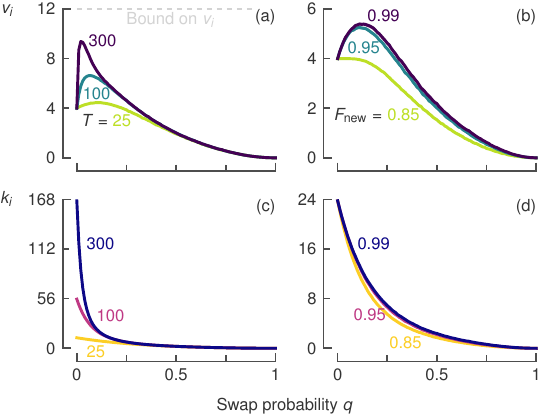}
    \caption{\textbf{Independent of the choice of network parameters, the optimal probability of attempting to swap~($q$) depends on the coherence time~$(T)$ and entanglement generation fidelity~($\Fgen$)}. (a, b) Virtual neighborhood size~($\vi$) and (c, d) virtual node degree~($\ki$) of a node in an infinite square lattice~($\di=4$) as a function of $q$. We vary (a, c) $T=25, 100, 300$ time steps (cutoff time $\tcut=3, 14, 42$ time steps) and (b, d) $\Fgen=0.85, 0.95, 0.99$ (maximum swap distance $M=1, 2, 3$); both colored from light to dark. Similar to Figure~\ref{fig:inf_sq} but for different network parameters, the optimal $q$ decreases with longer $T$ and increases for higher $\Fgen$, and~$\ki$ decreases monotonically as $q$ increases. Note that for long~$T$ and decreasing $q$, the maximum $\vi$ seems to approach the bound on $\vi$ (a). For a relatively short~$T=75$ ($\tcut=6$), the cutoff time quickly limits $\vi$ for increasing $\Fgen$ (b). We consider (a, c) $\Fgen=0.99$ ($M=2$) and (b, d) $T=75$ time steps ($\tcut = 6$ time steps). We assume that nodes generate entanglement and execute swaps probabilistically ($\pgen=0.99$, $\ps = 0.5$) and that nodes require a minimum link fidelity $\Fmin = 0.8$. Results obtained using network simulations and Monte Carlo sampling with $N=10^4$ realizations per sample, presented with an error band of $\pm 6s/\sqrt{N}$ (generally smaller than the line width), where $s$ is the sample standard deviation.}
    \label{fig:inf_sq_ext_data}
\end{figure}

\begin{figure}[H]
    \centering
    \includegraphics{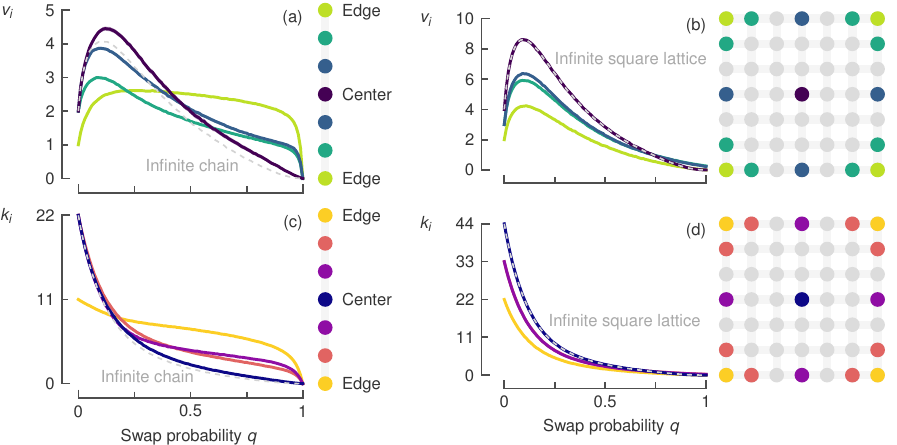}
    \caption{\textbf{Similar to the virtual neighborhood size~($\vi$), the behavior of the virtual node degree~($\ki$) as a function of the swap attempt probability~($q$) strongly depends on a node's location in a finite chain but, in a finite square lattice, it behaves qualitatively the same for all nodes.} Performance metrics (a, b) $\vi$ and (c, d) $\ki$ as a function of $q$ for nodes near and far from the network boundary (colored from light to dark) in a finite (a, c) chain~($\di=2$) and (b, d) square lattice~($\di=4$). We consider a coherence time $T=50$ time steps (cutoff time $\tcut=11$ time steps) and an entanglement generation fidelity $\Fgen=0.9$  (maximum swap distance $M=3$). We assume that nodes generate entanglement and execute swaps deterministically ($\pgen, \ps = 1$) and that nodes require a minimum link fidelity $\Fmin = \frac{1}{2}$. Results obtained with network simulations and Monte Carlo sampling with $N=10^4$ realizations per sample, presented with an error band of $\pm 6s/\sqrt{N}$ (generally smaller than the line width), where $s$ is the sample standard deviation.}
    \label{fig:fin_networks_inc_ki}
\end{figure}

\noindent In this Appendix, we provide additional data to the results presented in Section~\ref{sec:results}. In particular, we present both the virtual neighborhood size~$\vi$ and virtual node degree~$\ki$ for all the infinite regular topologies ($\di=2, 3, 4, 6$) for varying coherence times $T$ (Figure~\ref{fig:inf_networks_time_cutoff}) and for varying entanglement generation fidelity $\Fgen$ (Figure~\ref{fig:inf_networks_max_swap_dist}). Additionally, we present the performance metrics in an infinite square-lattice network for varying $T$ and $\Fgen$ with different network parameters (Figure~\ref{fig:inf_sq_ext_data}). Lastly, we present $\ki$ in addition to $\vi$ for finite chains and finite 
square lattices (Figure~\ref{fig:fin_networks_inc_ki}).

For increasing physical node degrees $\di$, the maximum virtual neighborhood size $\vi$ also increases (Figures~\ref{fig:inf_networks_time_cutoff} and~\ref{fig:inf_networks_max_swap_dist}). For example, increasing $d=2$ to $d=3$ increases the maximum value of $\vi$ by more than the ratio of node degrees~(3/2). We note that, for increasing maximum swap distance $M$, the bound on $\vi$ grows quicker in networks with $d = 3$ than those with $d = 2$
(see Appendix~\ref{app:sampling}, Table~\ref{tab:metric_bounds}). For example, increasing $M=1\to M=2$ (and assuming sufficiently large $\tcut$ such that $\vi$ is not bounded by the cutoff time), the bound on $\vi$ increases from $2\to 4$ when $d=2$, and from $3\to 9$ when $d=3$. Increasing the physical node degree to $d = 4$ and $d=6$ still increases the maximum $\vi$, but the increase relative to the ratio of physical node degrees (compared to $\di=2$) diminishes.

The performance metrics show the same qualitative behavior for different combinations of network parameters. For example, using probabilistic entanglement generation and execution of swaps ($\pgen=0.99, \ps = 0.5$; in contrast to deterministic generation and execution we previously assumed) and a higher minimum required fidelity ($\Fmin=0.8)$, ($i$) the virtual neighborhood size~$\vi$ still increases with longer coherence time~$T$ and better entanglement generation fidelity~$\Fgen$, $(ii)$ the optimal swap probability~$q$---maximizing $\vi$---decreases for longer~$T$ and increases for higher~$\Fgen$, and ($iii$) the virtual node degree~$\ki$ decreases monotonically for increasing $q$, going from $\ki=\di\cdot\tcut$ ($q=0$) to $\ki=0$ ($q=1$) (Figure~\ref{fig:inf_sq_ext_data}; for more details on the influence of probabilistic entanglement generation and swap execution on the behavior of the performance metrics, see Appendix~\ref{app:network-model-details}, Figure~\ref{fig:prob_gen_swap}). Additionally, we note that for long~$T$ (and associated long $\tcut$), the maximum~$\vi$ seems to approach the bound on~$\vi$ as nodes attempt very few swaps (low~$q$). For a maximum swap distance $M=2$, $\vi$ is bounded by $4\min\big(\tcut, \frac{1}{2}M(M+1)\big)=4\cdot\frac{1}{2}\cdot 2\cdot 3= 12$ (Table~\ref{tab:metric_bounds}). Furthermore, when $T$, $\tcut$ are relatively short, increasing~$\Fgen$ results in a limited increase in the maximum~$\vi$.

\vspace{10mm}
Similar to the virtual neighborhood size~$\vi$, the effect of network boundaries on the virtual node degree~$\ki$ depends on the network topology (Figure~\ref{fig:fin_networks_inc_ki}). For the same reasons as explained in the main text for $\vi$, the behavior of $\ki$ depends on the node's distance to the edge of a chain. In contrast, the behavior of $\ki$ is qualitatively similar for all nodes in a finite square-lattice network---decreasing monotonically from $\ki=d_i\cdot \tcut$ ($q=0$) to $\ki=0$ ($q=1$).

Lastly, for more data, we refer to the Jupyter Notebooks in this project's GitHub repository~\cite{code} and to the thesis~\cite{talsma2023continuous} that uses the same CD protocol.